\documentclass[twocolumn]{article}
\usepackage{romannum}
\usepackage{amsmath}
\usepackage{amssymb}
\usepackage{graphicx}
\usepackage{booktabs}   
\usepackage{relsize}
\usepackage{verbatim}
\usepackage{float}

\newcommand\arcsec{\mbox{$^{\prime\prime}$}}%
\usepackage{ulem}
\date{}
\usepackage{xcolor}
\graphicspath{{figures/}}
\usepackage{subcaption}

\usepackage{cancel}

\usepackage[super,comma,sort&compress]{natbib}
\usepackage[switch]{lineno}
\begin{document}

 \pagenumbering{arabic}
%\onecolumn

\setcounter{page}{63}

%\begin{figure*}[hbt!] %\vspace*{-4cm}
%\centering
 %     \includegraphics[scale=1]{Nature Astronomy Vol 7 Iss 6}
 %   \end{figure*}
    
\twocolumn[\title{\vspace{-5cm}\textbf{Einstein rings modulated by Wavelike Dark Matter from anomalies in gravitationally lensed images }}

\author{Alfred Amruth$^{1}$, Tom Broadhurst$^{2,3,4}$, Jeremy Lim$^{1}$,
Masamune Oguri$^{5,6,7}$, George F.  Smoot$^{3,8,9,10}$, \\Jose M.  Diego$^{11}$, Enoch Leung$^{12}$, Razieh Emami$^{13}$, Juno Li$^{1}$, Tzihong Chiueh$^{14,15,16}$,\\ Hsi-Yu Schive$^{14,15,16,17}$,  Michael C.  H.  Yeung$^{1}$, Sung Kei Li$^{1}$}

\maketitle
$^{1}$\textit{\small Department of Physics, University of Hong Kong, Hong Kong, Hong Kong SAR}\\
$^{2}$\textit{\small Department of Theoretical Physics, University of the Basque Country, UPV/EHU, 48080 Bilbao, Spain}\\
$^{3}$\textit{\small Donostia International Physics Center (DIPC), 20018 Donostia, Spain}\\
$^{4}$\textit{\small Ikerbasque, Basque Foundation for Science, E-48011 Bilbao, Spain}\\
$^{5}$\textit{\small Center for Frontier Science, Chiba University, 1-33 Yayoi-cho,
Inage-ku, Chiba 263-8522, Japan}\\
$^{6}$\textit{\small Research Center for the Early Universe, University of Tokyo, Tokyo
113-0033, Japan}\\
$^{7}$\textit{\small Kavli Institute for the Physics and Mathematics of the Universe
(Kavli IPMU, WPI), University of Tokyo, Chiba 277-8582, Japan}\\
 % $^{8}$\textit{Institute for Advanced Study and Department of Physics, The Hong Kong University of Science and Technology, Hong Kong}\\
$^{8}$\textit{\small Physics Department, University of California at Berkeley, CA 94720, Emeritus}\\
$^{9}$\textit{\small WF Chao Foundation Professor, IAS, Hong Kong University of Science
and Technology, Clear Water Bay, Kowloon, 999077 Hong Kong}\\
$^{10}$\textit{\small Paris Centre for Cosmological Physics, APC, AstroParticule et Cosmologie,
Universit\'{e} de Paris, CNRS/IN2P3, CEA/lrfu, 10, rue Alice Domon et Leonie Duquet,  
75205 Paris CEDEX 13, France {\it emeritus}}\\
$^{11}$\textit{\small Instituto de F\'isica de Cantabria (CSIC-UC).  Universidad de Cantabria.  Avda.  Los Castros s/n.  39005, Santander, Spain.}\\
$^{12}$\textit{\small Department of Physics and Astronomy, Johns Hopkins University, 3400 North Charles Street, Baltimore, MD 21218, USA}\\
$^{13}$\textit{\small Center for Astrophysics $|$ Harvard \& Smithsonian, 60 Garden Street, Cambridge, MA 02138, USA}\\
$^{14}$\textit{\small Institute of Astrophysics, National Taiwan University, Taipei 10617, Taiwan}\\
$^{15}$\textit{\small Department of Physics, National Taiwan University, Taipei 10617, Taiwan}\\
$^{16}$\textit{\small Center for Theoretical Physics, National Taiwan University, Taipei 10617, Taiwan}\\
$^{17}$\textit{\small Physics Division, National Center for Theoretical Sciences, Taipei 10617, Taiwan}

\vspace{1cm}

Corresponding author : Alfred Amruth, amruthalfredo@yahoo.com\\ 

{\bf Unveiling the true nature of Dark Matter (DM), which manifests itself only through gravity, is one of the principal quests in physics. Leading candidates for DM are weakly interacting massive particles (WIMPs) or ultralight bosons (axions), at opposite extremes in mass scales, that have been postulated by competing theories to solve deficiencies in the Standard Model of particle physics. Whereas DM WIMPs behave like discrete particles ($\varrho$DM), quantum interference between DM axions is manifested as waves ($\psi$DM).  Here, we show that gravitational lensing leaves signatures in multiply-lensed images of background galaxies that reveal whether the foreground lensing galaxy inhabits a $\varrho$DM or $\psi$DM halo.  Whereas $\varrho$DM lens models leave well documented anomalies between the predicted and observed brightnesses and positions of multiply-lensed images, $\psi$DM lens models correctly predict the level of anomalies left over by $\varrho$DM lens models. More challengingly, when subjected to a battery of tests for reproducing the quadruply-lensed triplet images in the system HS\,0810+2554, $\psi$DM is able to reproduce all aspects of this system whereas $\varrho$DM often fails. The ability of $\psi$DM to resolve lensing anomalies even in demanding cases like HS\,0810+2554, together with its success in reproducing other astrophysical observations, tilt the balance toward new physics invoking axions.}]

A universe operating under the tenets of General Relativity requires Cold Dark Matter (CDM), a hypothetical particle (or set of particles) that moves slowly, does not interact with light, and interacts with ordinary matter primarily or solely through gravity. The existence of such a particle demands a theory that extends the Standard Model (SM) of particle physics, if not an entirely new theory that includes or unifies current physical theories.  One extension of the SM, classified under supersymmetric theories, predicts the existence of Weakly Interacting Massive Particles (WIMPs) having rest-mass energies $\gtrsim$\,$10 \rm \, GeV$ (Ref.\,1). The lightest among the stable WIMPs has long been heralded as the most likely candidate for CDM. Laboratory searches, however, have failed to detect WIMPs through direct-detection or in collider experiments \cite{cosine}. Cosmological simulations employing massive bodies to stand in for WIMPs have been highly successful at predicting the large-scale structure of the universe, but face enduring problems on galactic or sub-galactic scales ($\lesssim$\,$10 \rm \, kpc$), the best documented of which are the ``missing satellite" (along with the related ``too big to fail") and ``cusp versus core" problems \cite{bullock}. Whether appealing to baryonic physics can resolve all these issues remains a subject of debate \cite{sales}.

At the opposite extreme in mass to WIMPs are axions: a broad class of particles that first appeared as the Peccei-Quinn solution to charge-parity (C-P) violation in the SM \cite{marsh}, but also in supersymmetric theories as well as theories with extra dimensions such as string theory \cite{svrcek,arvanitaki,marsh}(which seeks to unify all the four fundamental forces). Having rest-mass energies $\ll$\,$1 \rm \, eV$, these ultralight particles have no quantum-mechanical spin and therefore constitute bosons. Laboratory searches for DM axions began in the 1980's (Ref.\,8), with progressive improvements over time toward lower rest-mass energies \cite{bartram} albeit still many orders of magnitude above the energy range inferred from recent astronomical observations as mentioned below. Early theoretical studies referred to axions as fuzzy DM owing to the importance of quantum mechanical effects on such particles at macroscopic scales \cite{hu, peebles, sikivie}. In 2014, the first cosmological simulation employing ultralight bosons as CDM (Ref.\,13) confirmed the anticipated rich non-linear structure owing to self-interference of the Schrodinger-Poisson wave function describing the mean field behaviour of these particles, the defining characteristics of which are:\
 (i) an increasing suppression of self-gravitating concentrations (referred to as halos, which on forming stars constitute visible galaxies) toward lower masses owing to quantum pressure, which furthermore imposes a low-mass cutoff well above the Jeans limit; (ii) a soliton core comprising a coherent standing wave at the center of every halo; and (iii) self-interfering waves that fully modulate the DM density throughout the halo on the de Broglie scale, as has been corroborated in subsequent simulations \cite{mocz,veltmaat,niemeyer,hui}. Christened wavelike DM or $\psi$DM to emphasise its behaviour on macroscopic scales as waves, the de Broglie wavelength, $\lambda_{\rm dB}$, is set by the boson mass, $m_{\psi}$, for a given halo mass, $M_h$, according to the relationship\cite{schive2}:
         \begin{equation}
        \lambda_{\rm dB}=  150\left(\frac{10^{-22} \rm \, eV}{m_\psi}\right)\left(\frac{M_h}{10^{12} \, M_\odot}\right)^{-1/3} \rm pc
          \label{eq:m_psi1}
         \end{equation}
For a given $m_\psi$, $\lambda_{\rm dB}$ varies slowly with $M_h$ (i.e., $\lambda_{\rm dB} \propto M_h^{-1/3}$) as the higher momentum associated with a  large $M_h$ implies a smaller $\lambda_{\rm dB}$.  To strike a clear contrast with the more familiar form of CDM that behaves on macroscopic scales as particles (such as WIMPs), we shall refer to the latter as $\varrho$DM.  As we briefly summarise next, the growing success of $\psi$DM in reproducing  a variety of astronomical observations that are problematic for $\varrho$DM has called attention to ultralight bosons as a credible candidate for CDM; all such studies\cite{schive1,broadhurst,demartino,hui,enoch,pozo,pozoarxiv,herrera} advocate $m_\psi$ of order $10^{-22} \rm \, eV$. Other studies, however, deem $\psi$DM incompatible with some of the same astronomical observations\cite{read,hayashi2,read2,hayashi3}, or with entirely different astronomical observations if having $m_\psi \lesssim 10^{-21} \rm \, eV$ (Refs.\,29-30). All these studies either compare the inferred matter density fluctuations with the predicted suppression of $\psi$DM halos at low masses, or the inferred mass profile of low-mass galaxies with a solitonic core in $\psi$DM halos.  

While cosmological simulations involving both $\varrho$DM and $\psi$DM predict the same large-scale structure for the universe, $\psi$DM naturally gives rise to characteristics observed among galaxies that pose as problems for $\varrho$DM.  For instance, $\varrho$DM halos are predicted to increase dramatically in abundance toward lower masses until the Jeans limit well below $1 \rm \, M_{\odot}$ (Ref.\,3), thus giving rise to the missing-satellite problem\cite{bullock} that in turn prompt appeals to baryonic physics\cite{sales}. By contrast, $\psi$DM halos are predicted to be increasingly suppressed below masses of $\sim$$10^{10} (m_\psi/10^{-22} {\rm \, eV})^{-4/3} \, M_\odot$ until a cutoff at $\sim$$10^7 (m_\psi/10^{-22} {\rm \, eV})^{-3/2} \, M_\odot$ (Ref.\,14), thus helping alleviate the missing-satellite problem. Furthermore, the suppression of relatively low-mass $\psi$DM halos in the early universe provides an explanation for the apparent turnover in the abundance of galaxies toward lower luminosities at high redshifts\cite{enoch} (large cosmological distances). $\psi$DM having $m_\psi \lesssim 10^{-21} \rm \, eV$, however, appears to have difficulty in reproducing the density field at high redshifts as traced by the Ly-$\alpha$ forest (Ref.\,31).

For $m_{\psi} \sim 10^{-22} \rm \, eV$, $\lambda_{\rm dB}$ ranges from $\sim$100\,pc in massive galaxies ($M_h \sim 10^{11}$--$10^{12} \, M_\odot$) to $\sim$1\,kpc in dwarf galaxies ($M_h \sim 10^{9} \, M_\odot$), thus imposing a sizeable $\psi$DM solitonic core in, especially, low-mass galaxies. The latter offers an explanation for low-mass galaxies without cusps, potentially alleviating the ``cusp versus core'' problem in $\varrho$DM.  Indeed, several recent studies find that $\psi$DM halos featuring solitonic cores corresponding to $m_{\psi} \sim 10^{-22} \rm \, eV$ can reproduce the stellar velocity dispersions of dwarf galaxies\cite{broadhurst,pozo,pozoarxiv}, as well as their profiles in star counts\cite{pozoarxiv}. On the other hand, other analyses also based on stellar kinematics and which include some of the same dwarf galaxies find consistency with a CDM cusp in a number of cases\cite{read,hayashi2,read2,hayashi3} as predicted by $\varrho$DM. 

Here, we explore observational consequences for the third defining feature of $\psi$DM that provides the clearest signature of its quantum behaviour on macroscopic scales:\ the pervasive fluctuations in density ranging between zero (owing to completely destructive interference) and twice the local mean density  (owing to perfectly constructive interference) on a characteristic scale of $\lambda_{\rm dB}$. As we shall show, such density fluctuations in galactic halos can be revealed through their effects on gravitationally-lensed images of background galaxies, as hinted at in preliminary work by Ref.\,33. Indeed, lens models based on smoothly-varying density profiles as are motivated by $\varrho$DM commonly leave differences between the predicted and observed brightnesses \cite{keeton,nierenberg,goldberg,kochanek,shajib,xu} as well as, when observed at a sufficiently high angular resolution, positions of multiply-lensed images\cite{biggs,spingola,hartley}. These differences, well documented over the past 20\,years, are referred to as brightness (or flux) and position anomalies respectively. Populating massive halos with lower-mass halos -- referred to in this context as sub-halos, which when visible in starlight constitute satellite galaxies -- have had mixed success in resolving brightness anomalies\cite{amara,xu}, and have not yet addressed position anomalies let alone both anomalies simultaneously.  

In the following, we shall demonstrate that galactic $\psi$DM halos will inevitably leave both brightness and position anomalies at the observed levels -- provided that $m_{\phi}$ is of order $10^{-22} \rm \, eV$ -- if lens models are constructed for these halos based on smoothly-varying density profiles.  In making these comparisons, we adopt identical analytical functions for the global density profiles of both $\varrho$DM and $\psi$DM halos (the latter as smoothed over many $\lambda_{\rm dB}$) of the types most commonly adopted in the literature.  To highlight differences in lensed images generated solely by DM manifested macroscopically as waves rather than particles, we do not consider the role of sub-halos, if any, in producing brightness or position anomalies.  As discussed in Methods, sub-halos have very different physical properties than the pervasive over-dense (relative to the local mean) fluctuations in $\psi$DM.  Moreover, there is no counterpart in $\varrho$DM for the equally pervasive under-dense fluctuations in $\psi$DM.  Also crucially, sub-halos are much less abundant in $\psi$DM compared with $\varrho$DM as explained above.  We then show that $\psi$DM halos, unlike models based on $\varrho$DM halos, can actually reproduce the observed brightnesses and positions of the quadruply-lensed triplet images in the particularly well-observed system HS\,0810+2554 (Ref.\,37).  

  \begin{figure*}[hbt!] %\vspace*{-4cm}
         \renewcommand{\figurename}{Figure} 
       \hspace{-.4cm}
        \vspace{.4cm}
      \includegraphics[scale=0.34]{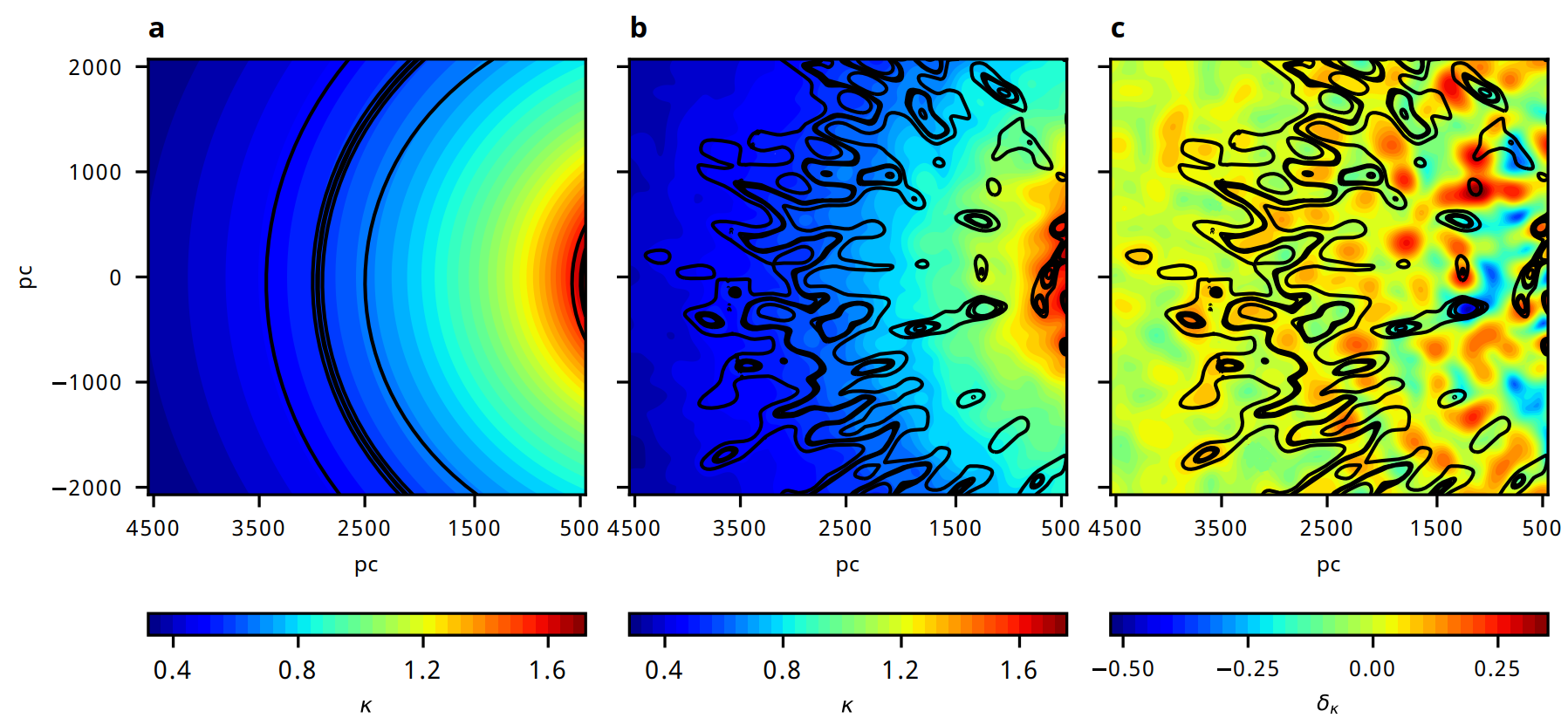}
      \caption{{\bf Iso-magnification contours of $\psi$DM versus $\varrho$DM halos.}  Iso-magnification contours for gravitational lensing by a halo at $z_l = 0.89$ imposed upon point sources at $z_s = 1.51$.  Contour levels are plotted at $\mu = 10$, 100, and 1000, where a thicker contour indicates the critical curve.  The $\varrho$DM halo has a NFW profile with parameters as listed in Extended Data Table\,\ref{tab:profiles}; the $\psi$DM halo has the same global profile, but imprinted onto which is a GRF having $\lambda_{\rm dB} = 180 \rm \, pc$ and a standard deviation, $\sigma_{\kappa}$, that varies with projected radius from the halo centre as plotted for a NFW profile in Extended Data Fig.\,\ref{fig:nfwvar}. The surface mass densities of these halos are expressed in terms of their convergence, $\kappa$ (see Methods).  {\bf a,} Smooth iso-magnification contours reflecting the smooth-varying surface mass density of the $\varrho$DM halo.  The critical curve (locus of highest magnification) can be easily recognised as a near-circular Einstein ring.  {\bf b,} Perturbed iso-magnification contours, including islands of high magnification, reflecting fluctuations in the surface mass density of a $\psi$DM halo.  {\bf c,} Same as {\bf b}, but showing only the GRF imprinted onto the $\varrho$DM halo in {\bf a} to create the corresponding $\psi$DM halo in {\bf b}. Perturbations in the iso-magnification contours can be seen to be directly related to fluctuations in the convergence, $\delta\kappa$ (see Methods), created by the GRF.  To highlight differences caused solely by quantum interference in $\psi$DM halos even when they have the same global profile as $\varrho$DM halos, we do not consider the uncertain presence of sub-halos throughout this article.}      

      \label{fig:kappa}
    \end{figure*}

\section*{Gravitational Lensing by $\psi$DM versus $\varrho$DM}
\vspace{-0.3cm}
Gravitational lensing by a foreground galaxy magnifies in both size and brightness (by preserving the surface brightness) background galaxies, but at the same time distorts the images of these galaxies owing to variations in lensing magnification across the image.  Where multiple images of the same background galaxy are produced, thus revealing gravitationally-lensed systems, none of the image positions correspond to the actual position of the background galaxy.  In this situation, the positions and brightnesses of the multiply-lensed images at known redshifts inform the projected surface (i.e., column) mass density of the foreground lensing galaxy as projected onto the sky.  Inaccuracies or missing ingredients in lens models thus constructed are revealed by their inability to exactly reproduce the observed positions and/or brightnesses of the multiply-lensed images.

    \begin{figure*}[htb!] 
\vspace*{-3cm}
\centering
%\hspace*{-2cm}
\vspace*{-0.2cm}
\includegraphics[scale=0.55]{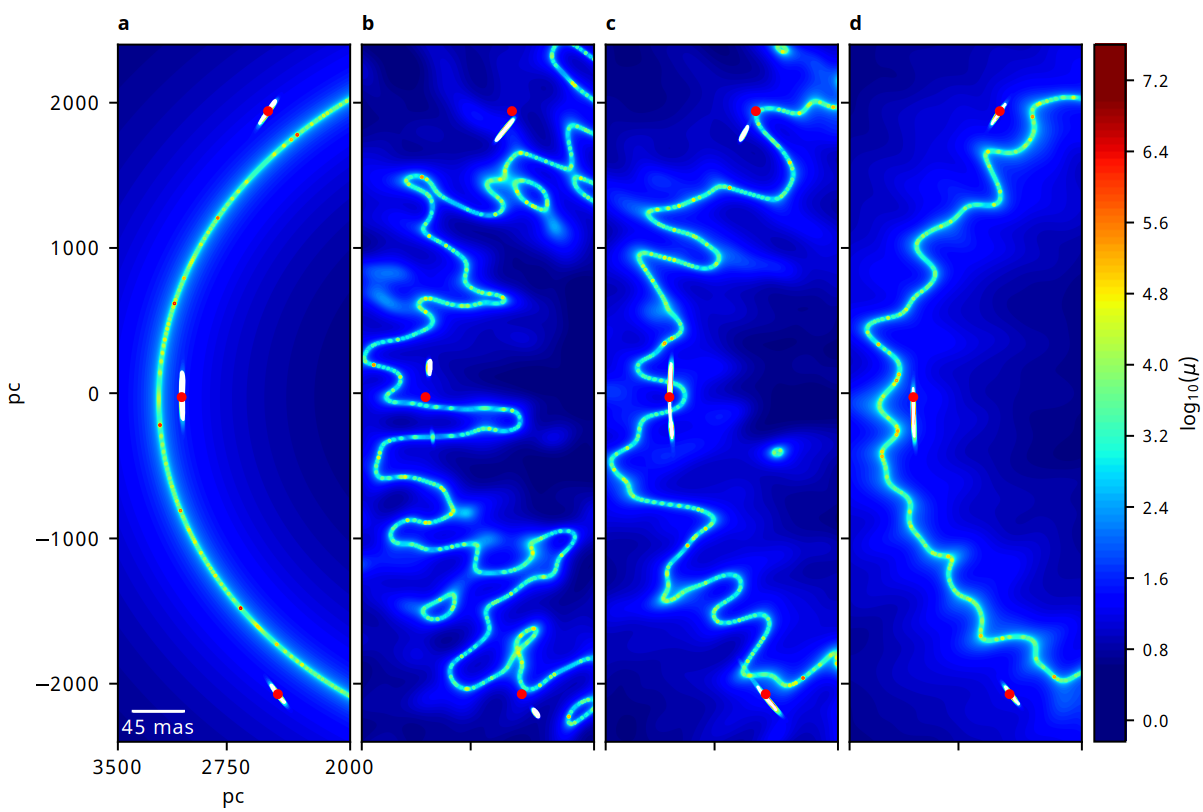}
\caption{{\bf Lensing Magnification of $\psi$DM versus $\varrho$DM halos.}  Rainbow colors indicating lensing magnification, $\mu$, imposed onto a source at $z_s = 1.51$ by a halo at $z_l = 0.89$ comprising either $\varrho$DM or $\psi$DM, both of which have the same global NFW profile (with parameters as listed in Extended Data Table\,\ref{tab:profiles}).   Physical scales are plotted at the redshift of the halo.   {\bf a,} Halo with a smoothly-varying density profile as is characteristic of $\varrho$DM.  {\bf b--d,} Halos onto which different GRF realisations have been imprinted onto the $\varrho$DM halo in {\bf a} to mimic $\psi$DM halos having $\lambda_{\rm dB} = 180 \rm \, pc$ (corresponding to $m_\psi = 1 \times 10^{-22} \rm \, eV$).  Fluctuations in lensing magnification by $\psi$DM decrease from {\bf b} to {\bf d} as the proportion of smoothly-distributed baryons within the Einstein ring increase. White arclets are lensed images of a compact circular source located at a fixed position near the cusp of a caustic (see Extended Data Fig.\,\ref{fig:imgcfg}). Their lengths are proportional to their lensing magnifications, which are visibly different for the corresponding arclets generated by the different halos.   Red dots correspond to the intensity-weighted centroids of the arclets in $\bf a$ as lensed by the $\varrho$DM halo, and are plotted repeatedly in $\bf b$--$\bf d$ to make clear the different positions of the lensed images generated by the $\psi$DM halos having the same global profile. These differences give rise to brightness and position anomalies when a model $\varrho$DM lens ({\bf a}) is used to reproduce the multiply-lensed images generated by $\psi$DM halos ({\bf b--d}), as quantified in Fig.\,\ref{fig:qso2}.}
\label{fig:qso}
\end{figure*}

Multiply-lensed images most commonly appear near the critical curve of the foreground lensing galaxy, where the lensing magnification is highest.  In situations where the foreground and background galaxies are in (near-)perfect alignment, the multiply-lensed images blend into an Einstein ring (which therefore corresponds to the critical curve of the lensing galaxy at the redshift of the background lensed galaxy).  Fig.\,1 (see construction in Methods) shows the surface mass density of a foreground lensing galaxy immersed in a $\varrho$DM (panel $a$) or $\psi$DM (panel $b$) halo, along with its corresponding critical curve.  Both halos have the same global density profile as described by a Navarro-Frenk-White (NFW) profile having the parameters listed in Extended Data Table\,1; in the case of the $\psi$DM halo, the global density profile corresponds to that smoothed over many $\lambda_{\rm dB}$.  To the $\psi$DM halo, we add a Gaussian random field (GRF) to represent, in projection, the pervasive density fluctuations of wavelike dark matter owing to quantum interference (see Methods). The redshift and mass of the lensing galaxy along with the radius of its Einstein ring (hereafter Einstein radius), as well as the redshift of the background lensed galaxy, match the corresponding parameters for the system HS\,0810+2554.  Whereas the $\varrho$DM halo has a smooth and nearly circular critical curve (panel $a$), the $\psi$DM halo exhibits a highly-perturbed critical curve along with isolated critical lensing islands (panel $b$) that are directly related to fluctuations in the surface mass density of this halo (panel $c$).  As might be anticipated, such fluctuations can lead to perturbations in both the brightnesses and positions of lensed images as we shall next demonstrate.

The lensing magnification generated by the same model halos as in Fig.\,1 are shown in Fig.\,2$a$ for $\varrho$DM and Fig.\,2$b$--$d$ for $\psi$DM (see construction in Methods).  To emphasise the indeterminate nature of quantum interference in an actual $\psi$DM halo, the $\psi$DM halos in Fig.\,2$b$--$d$ have different patterns in their random density fluctuations and therefore correspondingly different patterns in their critical curves.  These halos also have different mass fractions in stars and gas (collectively referred to as baryons): the incorporation of smoothly-distributed baryons is to dampen modulations in the 3-dimensional mass density field of a $\psi$DM halo, and therefore fluctuations in its corresponding projected 2-dimensional surface mass density field.  The damping introduced by smoothly-distributed baryons decreases away from the halo centre for a more centrally concentrated distribution of baryons than $\psi$DM.  Given the relatively short distances from the critical curve involved for the multiply-lensed images considered in Fig.\,2 as described next, for simplicity we damp the surface mass density fluctuations by constant factors of 20\%, 50\%, and 80\% in  Fig.\,2$b$--$d$ and consequently also the level of local perturbations in their lensing magnification; the corresponding fractional baryonic masses within the Einstein radius are summarised in Methods Data Table\,2 for different global density profiles for these baryons.

To visualise gravitational lensing by the model $\varrho$DM and $\psi$DM halos of Fig.\,2, we consider a compact and uniformly bright circular source located near the cusp of a caustic (projection of the critical curve onto the source plane as shown by the illustration in Extended Data Fig.\,2). The multiply-lensed images thus produced are shown as white arclets in Fig.\,2, for which only the highly-magnified triplet straddling the critical curve is shown -- these being the images normally considered in studies of brightness and position anomalies for cusp configurations. By comparison with the multiply-lensed images generated by the $\varrho$DM halo, those generated by the $\psi$DM halos have obviously different (total) brightnesses and (centroid) positions.

\begin{figure*}[htb!] 
\vspace*{-3cm}
\centering
\hspace*{-0.5cm}
\includegraphics[scale=0.55]{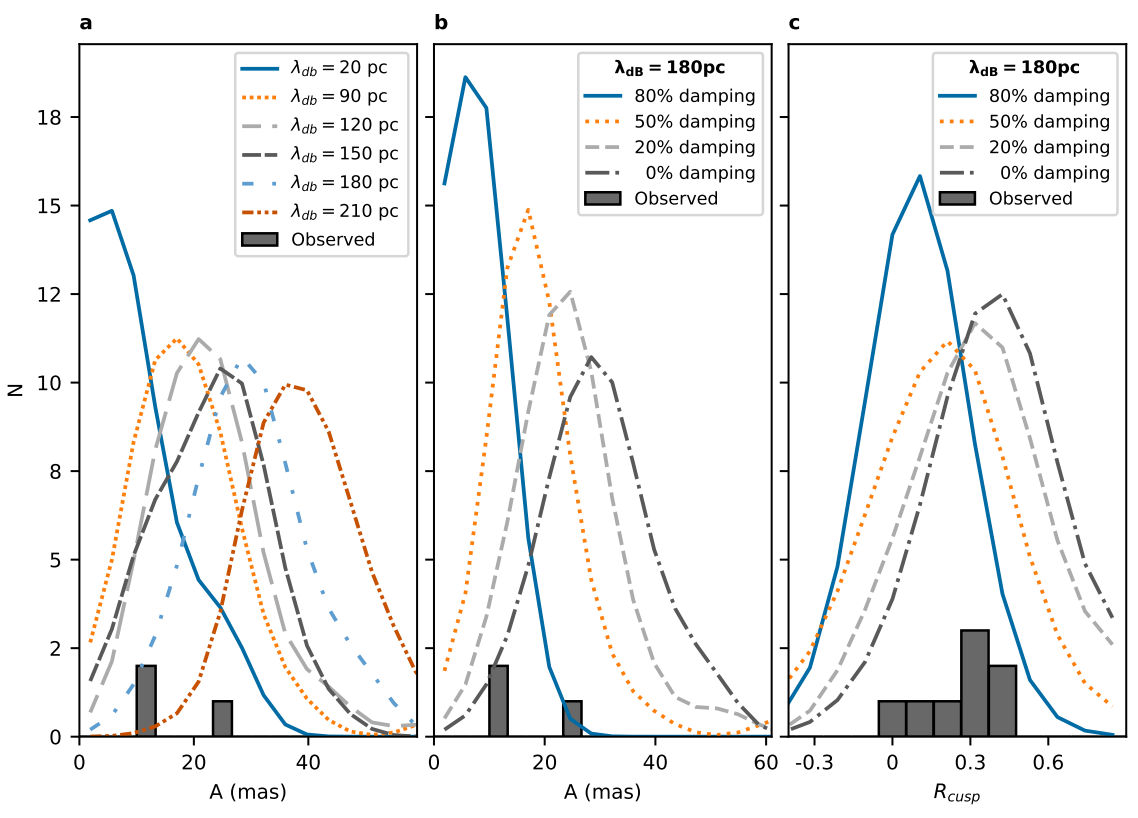}
\caption{{\bf Position and brightness anomalies.}  Probability distribution of position, $A$, and brightness, $R_{\rm cusp}$, anomalies (see text for how these parameters are defined) left when using the $\varrho$DM halo of Fig.\,\ref{fig:qso}$a$ to reproduce multiply-lensed images generated by $\psi$DM halos having the same global profile as this $\varrho$DM halo (examples shown in Fig.\,\ref{fig:qso}$b$--$d$).  {\bf a,} Different curves are position anomalies left for pure $\psi$DM halos (i.e., no baryons) having different $\lambda_{\rm dB}$ spanning the range 20--210\,pc, corresponding to the range in $m_{\psi}$ of (0.8--$8) \times 10^{-22} \rm \, eV$.  {\bf b,} Same as {\bf a} except now for $\psi$DM halos having a fixed $\lambda_{\rm dB} = 180 \rm \, pc$ but different fractional baryonic masses (see Extended Data Table\,\ref{tab:sersic}), resulting in damping of the fluctuations in their column mass density by different amounts. {\bf c,} Same as {\bf b}, except now for brightness anomalies.  The histograms in {\bf a}--{\bf b} show the reported position anomalies for the system HS\,0810+2554 (Ref.\,40) , whereas the histogram in {\bf c} shows the reported brightness anomalies for eight separate multiply-lensed QSOs\cite{nierenberg} including also HS\,0810+2554 (Ref.\,37) all based on observations with the HST. The predicted position and brightness anomalies are in broad agreement with those observed for $m_{\psi}$ of order $10^{-22} \rm \, eV$.}
 \label{fig:qso2}
\end{figure*}

\section*{Position and Brightness Anomalies}
\vspace{-0.3cm}
Fig.\,2 demonstrates that even a smooth lens model that accurately captures the global profile of $\psi$DM halos -- the $\varrho$DM halo shown in the same figure -- leave both position and brightness anomalies in multiply-lensed images owing to random density fluctuations in $\psi$DM halos. We now quantify the general level of position and brightness anomalies inevitably left by smooth lens models based on $\varrho$DM when constructed to reproduce multiply-lensed images generated by actual $\psi$DM halos.  For any given image configuration, the standard deviation in positional differences between lensed images thus produced provides a conventional measure of the positional anomaly.  That is, given a position $\mathbf{x}_i^{\psi {\rm DM}}$ for the (observed) $i$'th lensed image in the $\psi$DM lens (e.g., centroid of white arcs in Fig.\,2$b$--$d$), and a corresponding position $\mathbf{x}_i^{\rm \varrho DM}$ for the (predicted) $i$'th lensed image in the $\varrho$DM lens (i.e., centroids of white arcs in Fig.\,2\textit{a} as indicated by the red points in all panels), the position anomaly is given by $A = \left[\frac{1}{3}\sum_{i=1}^{3} (\mathbf{x}_i^{\psi {\rm DM}} - \mathbf{x}_i^{\rm \varrho DM})^2\right]^{1/2}$ for the highly magnified triplet images of a cusp configuration.  

Fig.\,3 shows the distribution in $A$ derived by comparing the multiply-lensed images generated by the model $\varrho$DM lens of Fig.\,2\textit{a} with those generated by pure $\psi$DM halos (i.e., no baryonic content) having the same global density profile (see Methods for its construction). The different colour curves in Fig.\,3\textit{a} correspond to different $\lambda_{\rm dB}$ depending on the selected $m_\psi$ (ranging over an order of magnitude around $10^{-22} \rm \, eV$): for a given $\lambda_{\rm dB}$, a range of position anomalies are possible owing to the indeterminate nature of quantum interference in actual $\psi$DM halos.  As would be naturally expected, the overall range in $A$ and its median value increases with increasing $\lambda_{\rm dB}$ and therefore decreasing values of $m_\psi$.  Fig.\,3$b$ shows the corresponding situation at a fixed $\lambda_{\rm dB} = 180 \rm \, pc$ (for $m_\psi = 1 \times 10^{-22} \rm \, eV$) but with the fluctuations in surface mass density of the $\psi$DM halo damped by different amounts according to the fractional baryonic mass within the Einstein radius (see Extended Data Table 2). Both the overall range in $A$ and its median value decreases with increasing baryonic content, as is the case shown in Fig.\,2$b$--$d$.  

Over the range of boson masses and baryonic content considered for the $\psi$DM halos, the predicted level of positional anomaly is of order 10 milliarseconds. Such small positional anomalies are not appreciable in optical observations even with the Hubble Space Telescope (HST), but noticeable in radio observations using Very Long Baseline Interferometry (VLBI): indeed, positional anomalies of this magnitude have been reported in VLBI observations of HS\,0810+255419 (Ref.\,40), MG J0751+271618 (Ref.\,41) and CLASS B0128+43717 (Ref.\,42). To compare our model predictions with actual observations, the histograms in Fig.\,3\textit{a}-\textit{b} indicate the positional anomalies left by the $\varrho$DM lens model constructed by Ref.\,40 for the highly-magnified triplet images straddling the critical curve of the quadruply-lensed system HS\,0810+2554 (recall that Figs.\,1--3 are constructed based on physical parameters appropriate for this system). Although the lensing configuration of this system is between a cusp and a fold thus requiring more exacting model calculations as are made below, the model predictions are in broad agreement with the measured position anomalies for $m_\psi$ of order $10^{-22} \rm \, eV$.

Fig.\,3\textit{c} shows the corresponding distribution in brightness anomaly, which is conventionally defined for a cusp configuration in terms of the magnifications  $\mu_1$, $\mu_2$, and $\mu_3$ for each of the triplet images formed near the critical curve by $R_{\rm cusp} =\frac{\mu_1 + \mu_2+ \mu_3}{|\mu_1| +|\mu_2| +|\mu_3|}$. 

Images formed outside the critical curve have positive parity (i.e., $\mu_1 > 0$ and $\mu_3 > 0$ for the configuration in Fig.\,2\textit{a}), whereas those that form inside the critical curve have negative parity ($\mu_2 < 0$). Studies have shown that small-scale structure has a higher probability of suppressing negative parity images than amplifying positive parity images\cite{keeton,diego}, resulting in a higher probability for obtaining a positive value of $R_{\rm cusp}$.  For the model $\varrho$DM lens, $R_{\rm cusp}$ = 0.09.  By comparison, for the model $\psi$DM lenses having a fixed $\lambda_{\rm dB} = 180 \rm \, pc$ but different fractional baryonic mass within the Einstein ring, $R_{\rm cusp}$ has a characteristic peak between 0.2 and 0.4 that depends only relatively weakly (unlike for $A$) on the baryonic content.  We note that the level of brightness anomaly predicted in Fig.\,3\textit{c} is similar to that predicted by Ref.\,33 based on a pure $\psi$DM halo of similar mass, although they do not report the predicted level of position anomaly. The brightness anomalies reported for eight separate systems\cite{nierenberg} including HS\,0810+2554 (Ref.\,37), all based on observations with the HST, are indicated by the histogram in Fig.\,3\textit{c}. Although the lensing galaxies involved span a range of masses, thus requiring model $\psi$DM lenses having different $\lambda_{\rm dB}$ to be generated for exacting one-on-one comparisons, the predicted brightness anomalies span a range similar to those reported for these galaxies. 

  \begin{figure*}[t!]
  \vspace*{-3cm}
        %\hspace{1cm}
        \centering
      \includegraphics[scale=0.55]{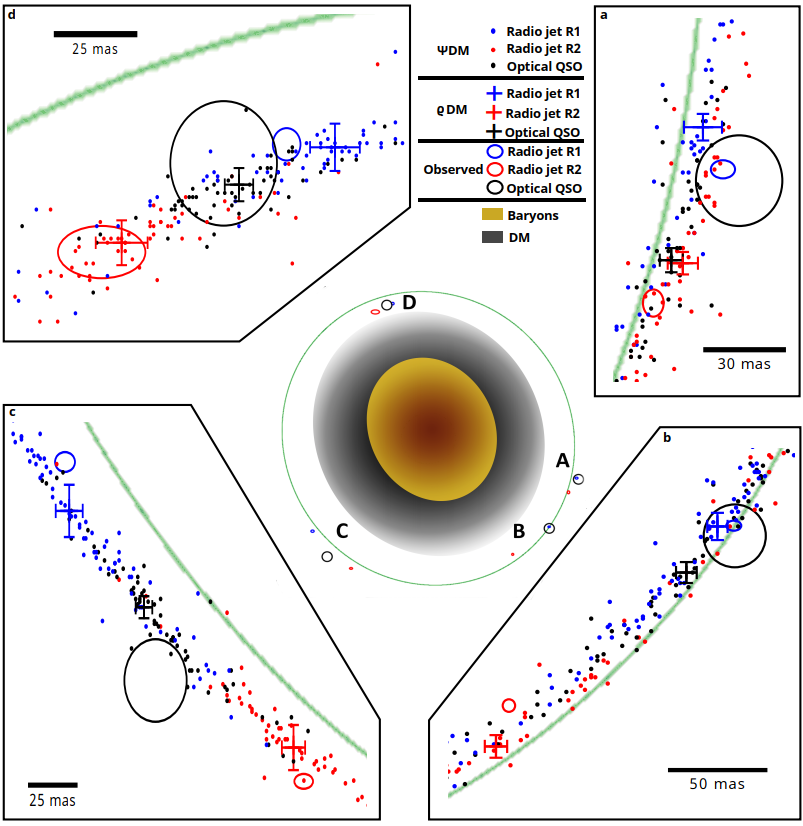}
      \caption{{\bf $\psi$DM versus $\varrho$DM model predictions for HS\,0810+2554.}  Predicted versus observed positions of an optical QSO (black symbols) and a pair of radio jets (red and blue symbols) at $z_s = 1.51$ that are quadruply-lensed (into images A--D) by an elliptical galaxy at  $z_l = 0.89$ (centred at origin of coordinate system).  Central panel is an overview of the entire system, where the green ellipse indicates the Einstein ring predicted by our best-fit $\varrho$DM lens model (see Methods).  Model galaxy coloured yellow for the baryonic component (size enclosing half its total mass) and grey for the $\varrho$DM component (size enclosing 15\% of its total mass).  {\bf a}--{\bf d,} Closeups on the individual sets of quadruply-lensed images labelled A--D, each comprising a single image of the QSO and two images for the pair of jets.  Observed image positions are indicated by circles or ellipses, each with a radius or semi-major/semi-minor axis of $3\sigma$ (where $\sigma$ is the measurement uncertainty) so as to encompass 99.7\% of all possible positions.  Crosses indicate the positions of the quadruply-lensed images predicted by our best-fit model $\varrho$DM lens, with arm lengths corresponding to $\pm 3 \sigma_{\rm tol}$ ($\sigma_{\rm tol}$ reflecting tolerances in the inferred parameters of the model $\varrho$DM lens; see Extended Data Fig. 6):\ these positions differ by much larger than the uncertainties for nearly all of the observed images.  Dots indicate predicted positions of the quadruply-lensed images based on 75 different GRF realisations having $\lambda_{\rm dB} = 180 \rm \, pc$ imprinted onto the model $\varrho$DM lens, mimicking a suite of $\psi$DM lenses all having the same global profile.  Baryons are smoothly distributed onto the $\psi$DM lenses to damp fluctuations in their surface mass densities (see text).  The positions of the lensed images predicted by the suite of model $\psi$DM lenses have uncertainties, owing to tolerances in the inferred global profile of the best-fit $\varrho$DM lens, similar to the corresponding crosses. Source positions as inferred from our by best-fit $\varrho$DM lens model is shown in Extended Data Fig.\,\ref{fig:glafic}$b$.}  
      \label{fig:hartley1}
    \end{figure*}

\section*{Reproducing the Gravitational-Lensed System HS 0810+2554}

The encouraging level of agreement between the predicted and measured level of anomalies demonstrated in Fig.\,3 motivates a more stringent test: evaluating whether $\psi$DM lenses can actually reproduce the observed positions and brightnesses of multiply-lensed images. For this test we consider the system HS\,0810+2554, which comprises a massive foreground elliptical galaxy that quadruply lenses a background galaxy featuring a:\ (i) quasi-stellar object (QSO) as imaged in the optical with the Hubble Space Telescope (HST)\cite{castles}; and (ii) pair of radio jets (presumably emanating from this QSO) as imaged with the European VLBI Network (EVN)\cite{hartley}.  Lensed images of each feature appear in all four sets of images labelled A--D in Fig.\,4.  The higher angular resolution attained with the EVN provides more precise image positions for the radio jets, as indicated by the four pairs of blue and red error ellipses in Fig.\,4, than the image positions measured by the HST for the optical QSO, as indicated by the four black error circles. On the other hand, whereas the brightness of the individual radio jets may not have been fully recovered owing to the lack of short baselines in the EVN observations, the brightnesses of the quadruply-lensed images of the optical QSO provide a more reliable measurement of the brightness anomaly. We do not make use of measurements of the QSO narrow-line emission made with the HST owing to blending of images A and B in the observation \cite{nierenberg}.

For lens modelling, the radio jets provide eight positional constraints whereas the optical QSO provides only four; both the radio and optical constraints cannot be used together owing to uncertainties in the registration between optical and radio references frames amounting to a few 10\,mas.

Using therefore the positions of the quadruply-lensed radio jets as constraints, we constructed different $\varrho$DM lens models with and without external shear (see Methods) to estimate the global radial density profile of the lensing galaxy, and settled on a best-fit model comprising an elliptical power-law profile without external shear (see parameters in Extended Data Table 1).  The resulting model-predicted image positions for the quadruply-lensed pair of radio jets are indicated by four pairs of red and blue crosses in Fig.\,4 (having arm len  gths reflecting tolerances in the model $\varrho$DM lens; see Methods), many of which are significantly different from their observed positions as indicated by the corresponding pairs of blue and red error ellipses.  The position anomalies thus left by our best-fit $\varrho$DM lens model are similar in level with those left by the $\varrho$DM lens model reported by Ref.\,40. To best match the observed image positions of the optical QSO, our $\varrho$DM lens model requires this QSO -- the presumed core of the radio jets -- to be closer to one jet than the other (see Extended Data Fig.\,5), as is typically observed for core-jet systems owing to stronger Doppler boosting of one jet than its opposing counterpart.  The image positions thus predicted by our $\varrho$DM lens model for the quadruply-lensed QSO are indicated by the four black crosses in Fig.\,4 (again having arm lengths reflecting tolerances in the model $\varrho$DM lens), once more leaving position anomalies for most of the lensed images just like in the $\varrho$DM lens model of Ref.\,40.

To construct the model $\psi$DM lens having the same global radial density profile as the model $\varrho$DM lens, we imprinted seventy-five different randomly generated patterns for its density fluctuations -- all having $\lambda_{\rm dB} = 180 \rm \, pc$ (corresponding to a boson mass of $1 \times 10^{-22} \rm \, eV$ given the halo mass) -- onto the model $\varrho$DM lens so as to reflect the indeterminate nature of quantum interference in an actual $\psi$DM halo (see Methods).  Our $\psi$DM model for the lensing galaxy includes a baryonic component that damps fluctuations in its surface mass density by 50\% around the critical curve by comparison with a pure $\psi$DM halo (see Extended Data Table 2 for the corresponding fractional mass in baryons depending on its global profile).  Fig.\,4 shows the resulting model-predicted positions for the pair of radio jets as indicated by the blue and red points, as well as for the optical core as indicated by the black points, for which seventy-five points are plotted for each lensed feature at each image set A--D.  Tolerances in the inferred parameters for the global radial density profile of the actual $\psi$DM halo give rise to uncertainties in the model-predicted positions of the lensed images, for which the sizes of the crosses in Fig.\,4 provide a gauge.  As can be seen, perturbations in the positions of the quadruply-lensed lensed images introduced by fluctuations in the surface mass density of the model $\psi$DM lens can bring the model-predicted positions to good agreement with the observed positions -- although finding a simultaneous one-to-one match for every lensed image is highly unlikely given the random nature of density modulations in a $\psi$DM halo (see Extended Data Fig.\,7).

Although the positional deflection imposed on a given lensed image is independent of any other given the random nature of density fluctuations in a $\psi$DM halo, an important test is to check whether the suite of model $\psi$DM lenses predicts a preferential angular separation between the different features in a given set of lensed images; this preferred angular separation, if it exists, would therefore also correspond to that most likely to be observed. Supplementary Fig.\,1 shows that there is indeed a preferred angular separation between the radio jets predicted by the suite of model $\psi$DM lenses:\ for the majority of jet pairs, this preferred angular separation is in close agreement with their observed angular separation. Supplementary Fig.\,2 demonstrates the ability of the suite of model $\psi$DM lenses to reproduce the observed ratio in intensities between the pair of radio jets in each of the four sets of lensed images. Finally, Supplementary Fig.\,3 shows the ability of the suite of model $\psi$DM lenses to reproduce the brightness anomaly left over by our $\varrho$DM lens model for the optical QSO.  The broad agreement found between model predictions and observations in all these tests demonstrates the predictive power of $\psi$DM for reproducing the multiply-lensed images of HS\,0810+2554. By contrast, this system continues to pose a challenge for $\varrho$DM: future efforts incorporating triaxial DM halos or baryonic structures with morphologies different from the DM halo, along with sub-halos, may be required to assess whether current lensing anomalies can be reconciled using $\varrho$DM lens models.

As mentioned earlier, to make clear the effects of pervasive density modulations in $\psi$DM halos on gravitationally-lensed images, we do not include any sub-halos around the lensing object. Not only are such density modulations much more pervasive and have entirely different properties than sub-halos (e.g., producing under-dense fluctuations as often as over-dense fluctuations), the latter is increasingly suppressed below masses of $\sim$$10^{10} \rm \, M_\odot$ (for $m_\psi \sim 10^{-22} \rm \, eV$) in the case of $\psi$DM halos. This suppression makes it even more difficult for sub-halos to account for brightness anomalies compared to their much more numerous counterparts around $\varrho$DM halos.  In a recent work \cite{laroche} that incorporates sub-halos around both $\varrho$DM and $\psi$DM halos, neither were found to be able to reproduce the brightness ratios between all possible pairs of lensed images in multiply-lensed QSOs, such that $\psi$DM halos having $m_\psi < 10^{-21.5} \rm \, eV$ can generate even more disparate brightness ratios than $\varrho$DM halos. Satisfying this stringent demand requires finding realisations that match all observed brightness ratios in a given system, a highly unlikely prospect given the multitude of ways in which sub-halos can be arranged around $\varrho$DM halos let alone both subhalos and density modulations around $\psi$DM halos; furthermore, that $\psi$DM can produce even larger disparities than $\varrho$DM is no surprise, owing to relatively strong but rare over- or under-densities that happen to coincide with a lensed image. In our work, we assess whether different $\psi$DM realisations can reproduce individual aspects of at least one of the lensed images (comprising multiple components) in HS\,0810+2554, and rely on the ability to reproduce the same aspect in all the lensed images among different realisations to infer the existence of at least one realisation that matches all such aspects of the system.

\begin{flushleft}
\textbf{\large Future Prospects} 
\end{flushleft}

The increasing astrophysical evidence for ultralight bosons with rest-mass energies of order $10^{-22} \rm \, eV$ (Refs.\,13,14,19-24) has propelled axions -- a class of particles well motivated by theories of new physics -- to the forefront as a candidate for CDM.  New observational consequences of $\psi$DM continue to be evaluated and subjected to astrophysical tests. Laboratory experiments to detect DM axions continue, and new experiments are being proposed and developed.  Laboratory experiments designed to detect WIMPs at sensitivities reaching the neutrino floor (as imposed by cosmic, terrestrial, and man-made neutrinos), slated for this decade, will provide a critical reckoning for the class of new physics that predict WIMPs -- and with it the viability of these particles as candidates for CDM. Crucially, determining whether $\varrho$DM or $\psi$DM better reproduces astrophysical observations will tilt the balance towards one of the two corresponding classes of theories for new physics.

%%%%%%%%%%% END OF MAIN TEXT %%%%%%%%%%%%%%%%%%%

\newpage

%%%%%%%%%%%%%%%%%%%%%%%%%%%%%%%%%%%%%%%%%%%%%%
%Methods Section
\newpage
        \begin{center}
\textbf{\Huge Methods}
    \end{center}
\section{\textbf{Halo Mass Density Fields}} \label{subsec:nfw}

   \begin{table*}[htb!]
       \renewcommand{\tablename}{Extended Data Table}
    \begin{tabular}{p{2.8cm}ccccccp{0.5cm}}
      \toprule
      \toprule
    {\centering Model}  & {\centering$z_l$ }& {\centering $z_s$} & {\centering Einstein Radius (\arcsec{})} & Ellipticity & Scale Radius (kpc) & {\centering $M_h (10^{11} M_\odot$)} & c \\ \hline
      NFW & 0.89 & 1.51 & 0.46 &  0.2 & 5 & 7& 9\\
            PL (index of 1.7) & 0.89 & 1.51 & 0.48 &  0.05 & - & - & -\\ 
            \bottomrule
    \end{tabular}
    \caption{Parameters for lens models. Note: $M_h$ is defined as virial mass and c is concentration parameter.}
                \label{tab:profiles}
  \end{table*}

\subsection{Global Radial Density Profile}
Figs.\,\ref{fig:kappa}--\ref{fig:qso2} are constructed based on a
%compare with the observations described in the main text that find both brightness and positional anomalies in multiply-lensed images when adopting halos having smoothly-varying density profiles for the foreground lensing galaxy, 
foreground lensing galaxy at a redshift of $z_{l}$ = 0.89 having a halo virial mass of $M_h = 7 \times 10^{11} M_\odot$, and a background lensed galaxy at a redshift of $z_{s} = 1.51$, so as to provide a specific match to the system HS\,0810+2554 (Ref.\,40).  We adopt a Navarro-Frenk-White (NFW) profile for the global radial density profile of the lensing galaxy, as is commonly used to model CDM halos.  A small ellipticity of $e = 0.2$ is imposed onto the halo such that its symmetry axis is aligned with the sky.  We selected a concentration parameter and scale radius for this NFW profile so as to give rise to an Einstein radius corresponding to that inferred for the lensing galaxy in HS\,0810+2554.  This model halo, the parameters for which are summarised in Extended Data Table\,\ref{tab:profiles}, stands in for a $\varrho$DM halo devoid of sub-halos so as to highlight brightness and position anomalies arising solely from the pervasive density fluctuations characterising $\psi$DM halos.  Fig.\,\ref{fig:hartley1} was constructed based on the same aforementioned redshifts for the foreground lensing and background lensed galaxy, as well as the same halo mass for the lensing galaxy, but with the global density profile of the lensing galaxy tuned to best fit the positions of the multiply-lensed images in HS\,0810+2554 as described in more detail later.

\subsection{Projection onto the Sky}
Owing to the large cosmic separations involved, gravitational lensing by a galaxy can be accurately treated as a thin lens -- corresponding to the column mass density of the galaxy as projected onto the sky.  Such a projection from three dimensions (3-D) to two dimensions (2-D) is straightforward for halos having a smoothly-varying radial density profile.  To compute the 2-D surface mass density of a $\psi$DM halo, however, we need to integrate along the line-of-sight through its 3-D mass density field that fluctuates randomly about the local mean owing to quantum interference.  The 3-D mass density field of $\psi$DM halos is inherently complex\cite{schive1}: over a spatial scale characterised by the de Broglie wavelength, $\lambda_{\rm dB}$ (see Eq.\,\ref{eq:m_psi1}), the density can fluctuate randomly between zero (corresponding to fully destructive interference) to twice (corresponding to fully constructive interference) the local average density.  As we do not know a priori the distribution in 3-D densities along a given sightline, the line-of-sight integration through the randomly fluctuating 3-D density field of a $\psi$DM halo is inherently indeterministic.  Nonetheless, if performed repeatedly for all possible outcomes for the 3-D density distribution along a particular sightline, this integration tends toward a Gaussian distribution regardless of the distribution in 3-D densities as dictated by the central limit theorem.  As we will show, the column mass density field of DM halos can therefore be approximated as a Gaussian random field (GRF) having also a characteristic scale of $\lambda_{\rm dB}$ imprinted onto the smoothly-varying (i.e., averaged over many $\lambda_{\rm dB}$ across the sky) global density profiles of these halos as projected onto the sky.  Given an analytical function for the global density profile of a $\psi$DM halo, the manner by which the variance of its associated GRF changes with projected radius from the halo centre can be computed in the manner described below.

Let the halo mass density be $\rho(\mathbf{r})$ at a 3-D position vector $\mathbf{r}$ from the center of a halo, around which the mean density (i.e., smoothed over many $\lambda_{\rm dB}$) as determined from the 3-D global density profile is $\langle \rho(\mathbf{r})\rangle$.  As mentioned above, $\rho(\mathbf{r})$ can fluctuate randomly between zero and 2$\langle \rho(\mathbf{r})\rangle$, such that at a given position $\mathbf{r}$ the fluctuation from the local mean is $\delta\rho = \rho(\mathbf{r}) - \langle \rho(\mathbf{r})\rangle$.  At a projected radius vector $\boldsymbol{\xi}$ from the center of the halo, integrating along the line of sight yields the column density $\Sigma(\boldsymbol{\xi})$.  Applying the central limit theorem over a small region around $\boldsymbol{\xi}$, the column density fluctuates randomly about the local mean column density, $\langle \Sigma(\boldsymbol{\xi}) \rangle$, with a Gaussian distribution.  The deviation in column density from the local mean, $\delta\Sigma = \Sigma(\boldsymbol{\xi}) - \langle \Sigma(\boldsymbol{\xi}) \rangle$, can be computed according to $\delta\Sigma =\lim_{n \to \infty} \sum^n_{i=1} \delta\rho_i\Delta z_i$, where $i$ denotes the many stochastic (random and independent) 3-D density fluctuations from the local mean, $\delta\rho_i$, encountered along the sightline, and $\Delta z_i$ is an interval equal to $\lambda_{\rm dB}$ in the line-of-sight direction $z$.  The variance in column density fluctuation, $\sigma^2_\Sigma(\boldsymbol{\xi})$, can be computed by taking the expectation value of the squared deviations:

\setlength{\abovedisplayskip}{2pt}
\setlength{\belowdisplayskip}{2pt}
\setlength{\abovedisplayshortskip}{2pt}
\setlength{\belowdisplayshortskip}{2pt}
\begin{align}  \label{EM1}
    \sigma^2_\Sigma(\boldsymbol{\xi}) &= \Big\langle\Big(\lim_{n \to \infty} \sum^n_{i=1}\delta\rho_i\Delta z_i\Big)^2\Big\rangle  
\end{align}

Now, by definition, the expectation value is obtained by multiplying a variable with its probability and then summing over all possible outcomes for this variable.  Owing to the stochastic nature of $\psi$DM fluctuations, each fluctuation draws its density from its own unique probability density function (PDF), $P_{\delta\rho,i}$; to obtain the expectation value, we therefore need to multiply each fluctuation with its respective PDF.  The product of all the PDFs in sequence (i.e., multiplicative version of the summation sign) is denoted by the symbol $\prod_{i=1}^n$, for which the symbol $d^n(\boldsymbol{\delta\rho})$ (the integration measure) indicates an integration over all the fluctuations (from $i = 1$ to $n$). The variance in column density fluctuations is therefore:

\begin{align}   \label{EM2}
 \sigma^2_\Sigma(\boldsymbol{\xi})&= \lim_{n \to \infty}\int_n d^n(\boldsymbol{\delta\rho}) \Big[\sum^n_{i=1}(\delta\rho_i\Delta z_i)^2 +  \nonumber\\
    & \sum^n_{i=1}\sum^n_{j\neq i} \cancelto{0}{\delta\rho_i\delta\rho_j\Delta z_i} \Delta z_j \Big]\prod^n_{i=1} {P_{\delta\rho,i}}  .
    \end{align}

\noindent Because the fluctuations are independent from each other, there is no correlation between fluctuations and consequently the cross-term is 0.  

Expanding Eq.\,\ref{EM2}:

         \begin{align}   \label{EM3}
   \sigma^2_\Sigma(\boldsymbol{\xi}) &= \lim_{n \to \infty} \sum^n_{i=1}(\Delta z_i)^2 \Big(\int^\infty_{-\infty}\Big[(\delta\rho_i)^2 P_{\delta\rho,i} \Big] d(\delta\rho_i) \nonumber\\ &  \int_{n-1} \cancelto{1}{d^{n-1}(\boldsymbol{\delta\rho}) \Big(\prod^n_{j\neq i}P_{\delta\rho,j}}\Big)  \Big)
   \end{align}

The first integral, which picks out the PDF corresponding to the $i$-th fluctuation, is, by definition, the variance in $\delta\rho_i$ given its particular $P_{\delta\rho,i}$, and will henceforth be denoted as $\sigma^2_{\rho,i}$.  The second integral, which picks out the PDFs corresponding to all other fluctuations, evaluates to unity as it comprises an integral
over the entire PDF for each fluctuation.  Eq.\,\ref{EM2} can therefore be rewritten as:

   \begin{align}     \label{EM4}
      \sigma^2_\Sigma(\boldsymbol{\xi}) &= \lim_{n\to\infty}\sum^n_{i=1}\sigma^2_{\rho,i}(\Delta z_i)^2
   \end{align}

\vspace{0.3cm}
The summation as expressed by Eq.\,\ref{EM4} should be performed over intervals $\Delta z_i$ that are much smaller than the scale of variations in $\sigma^2_{\rho}$, that is $\Delta z << \sigma^2_\rho / |d\sigma^2_\rho/dz|$ as is the case when $\Delta z_i \sim \lambda_{\rm dB}$ or smaller.  In this situation, we can promote the discrete sum of Eq.\,\ref{EM4} to the continuous limit by treating $\sigma^2_{\rho,i} = (1/\Delta z_i)\int_{z_i}^{z_{i+1}}\sigma^2_{\rho} \, dz$; i.e., the discretised mass density variance, $\sigma^2_{\rho,i}$, can be interpreted as the average value of the continuous mass density variance, $\sigma^2_\rho$, within the $i$-th fluctuation. Applying this treatment,
\\

   \begin{align}    \label{EM5}
    \sigma^2_\Sigma(\boldsymbol{\xi}) &= \lim_{n\to\infty}\sum^n_{i=1}\Big[\frac{1}{\Delta z_i} \int_{z_i}^{z_{i+1}}\sigma_\rho^2 \ dz\Big] (\Delta z_i)^2
   \end{align}
\vspace{0.2cm}
\\
Substituting $\Delta z_i$ by $\lambda_{\rm dB}$,
   \begin{align}    \label{EM6}
   \sigma^2_\Sigma(\boldsymbol{\xi}) &= \lambda_{\rm dB} \Big(\lim_{n\to\infty}\sum^n_{i=1}\int_{z_i}^{z_{i+1}} \sigma^2_{\rho}\ dz \Big)  \nonumber  \\
   &= \lambda_{\rm dB} \int_{-\infty}^{\infty} \sigma^2_\rho(z,\boldsymbol{\xi})dz
\end{align}
\\
meaning that the variance in the column mass density can simply be written in terms of the variance in the 3-D mass density along a given sightline.  In Eq.\,\ref{EM6}, we have made explicit that $\sigma^2_\rho$ is a function of the 3-D position $\mathbf{r} \rightarrow (z,\boldsymbol{\xi})$.

As an example, let us consider a $\psi$DM halo having a global density profile described by a NFW profile.  If the 3-D density of this NFW profile at $\mathbf{r}$ is denoted as $\rho_{\rm smooth}(z,\boldsymbol{\xi})$, then the standard deviation in density at $\mathbf{r}$ for the corresponding $\psi$DM halo is $\sigma_{\rho}(z,\boldsymbol{\xi}) \sim \rho_{\rm smooth}(z,\boldsymbol{\xi})$ (as the 3-D density fluctuates between 0 and twice the local mean density).  The exact proportionality between $\sigma_{\rho}(z,\boldsymbol{\xi})$ and $\rho_{smooth}(z,\boldsymbol{\xi})$ is not of great concern because, in a real galaxy, $\sigma_{\rho}(z,\boldsymbol{\xi})$ is damped to varying degrees depending on the radial distribution of a smoothly-varying baryonic component as discussed below.  In this situation:
\\

\begin{align}   \label{EM7}
 \sigma^2_\Sigma(\boldsymbol{\xi}) &= \lambda_{\rm dB}\int_{-\infty}^{\infty} \sigma^2_\rho(z,\boldsymbol{\xi})dz \nonumber \\
 &\simeq \lambda_{\rm dB}\int_{-\infty}^{\infty} \rho_{\rm smooth}^2(z,\boldsymbol{\xi})dz \nonumber \\
&\simeq \begin{cases}
\mathlarger{\rho^2_{\rm o} r_s \lambda_{\rm dB}\Bigg[\frac{\pi}{x}- \frac{1}{(x^2-1)^3}\Bigg[\frac{6x^4-17x^2+26}{3}+}\\
\ \ \ \ \mathlarger{\frac{2x^6 -7x^4 +8x^2-8}{\sqrt{1-x^2}}\mathrm{sech}^{-1}(x) \Bigg]\Bigg]  ,~~  x<1,}\\ \\
\mathlarger{\rho^2_{\rm o} r_s \lambda_{\rm dB}\Bigg[\frac{\pi}{x}- \frac{1}{(x^2-1)^3}\Bigg[\frac{6x^4-17x^2+26}{3}+}\\
\ \ \ \ \mathlarger{\frac{2x^6 -7x^4 +8x^2-8}{\sqrt{x^2-1}}\sec^{-1}(x) \Bigg]\Bigg ]  ,~~ x>1.}
\end{cases}
\end{align}
\\
where $\rho_{\rm o}$ is a normalization factor ($\rho_{\rm smooth}\equiv \rho_{\rm o}/4$ at the scale radius, $r_s$), and $x \equiv \boldsymbol{\xi}/r_s$.  In the formalisation of gravitational lensing, we can define a dimensionless quantity known as the convergence, $\kappa$, which expresses the normalised column density as $\kappa = \Sigma(\boldsymbol{\xi}) / \Sigma_{\rm cr}$, where $\Sigma_{\rm cr}$ is the critical surface density.  The latter is related to the lensing geometry by $\Sigma_{\rm cr} = c^2D_s/(4\pi D_l D_{ls})$, where $D_s$ is the angular diameter distance to the lensed source, $D_l$ is the angular diameter distance to the lens, and $D_{ls}$ is the angular diameter distance between the lens and the lensed source. The deviation in column density from the local mean in terms of the convergence can be written as $\delta\kappa = \delta\Sigma/\Sigma_{\rm cr}$. In the same way, we can define a dimensionless quantity for the variance in column density fluctuations, $\sigma^2_\kappa(\boldsymbol{\xi}) \equiv \sigma^2_\Sigma(\boldsymbol{\xi})/\Sigma^2_{\rm cr}$.  A detailed calculation in Fourier space arrives at a similar form for the variance in column density of the $\psi$DM fluctuations\cite{kawai}.

    \begin{figure}[hbt!] %\vspace*{-4cm}
        \setcounter{figure}{0}
         \renewcommand{\figurename}{Extended Data Figure}      %\hspace{-1cm}
      \includegraphics[width=8cm,height=6cm]{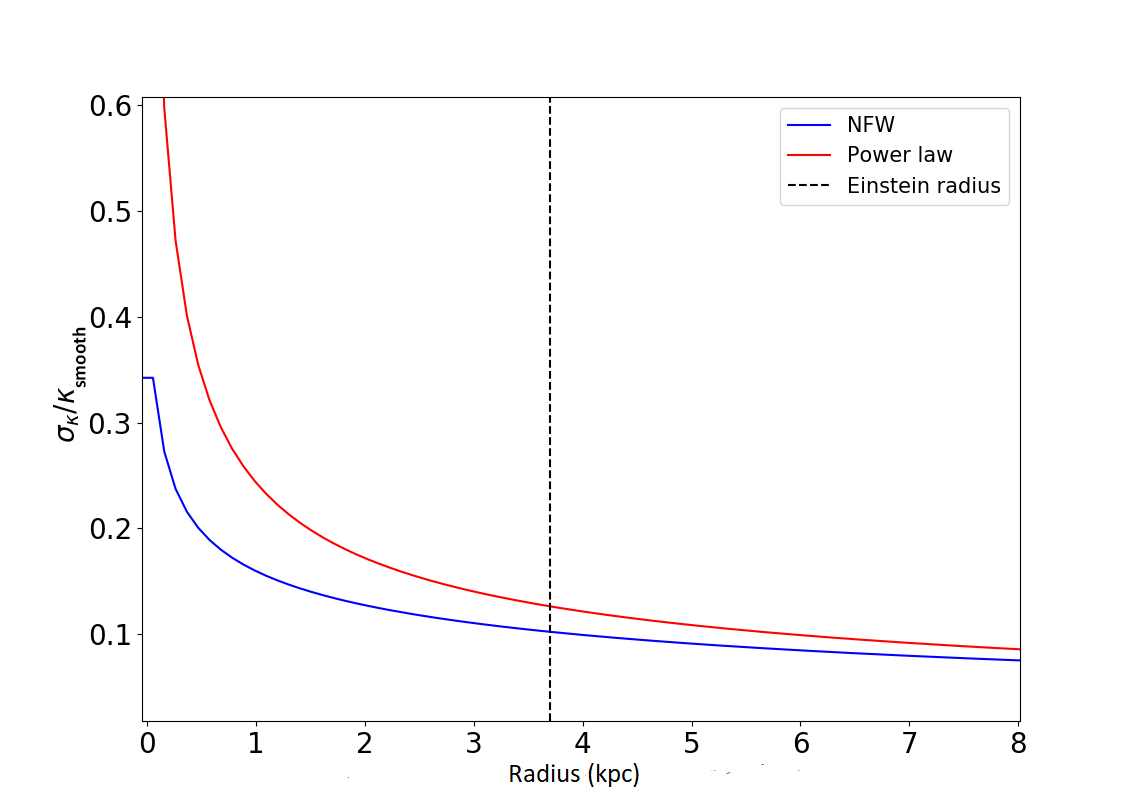}
      \caption{\textbf{Standard deviation of GRF}.  Projected radial dependence in $\sigma_\kappa(\boldsymbol{\xi}) \equiv \sigma_\Sigma(\boldsymbol{\xi})/\Sigma_{\rm cr}$ for the GRF relative to $\kappa_{\rm smooth} \equiv \Sigma_{\rm smooth}(\boldsymbol{\xi}) / \Sigma_{\rm cr}$ for an NFW (blue) global profile computed according to Eq.\,\ref{EM7}, and for a PL (red) global profile computed in the same manner.  Imprinting GRFs having an appropriate $\sigma_\kappa$ onto these profiles to generate model $\psi$DM lenses, fluctuations in the column mass density of these lenses diminish with increasing projected radius, $\boldsymbol{\xi}$, from the halo centre.  At the Einstein radius for these particular model lenses and the lensed source (see main text), the GRF has a standard deviation of $\sim$10\%-15\% the local mean column mass density of both these profiles.}
      \label{fig:nfwvar}
    \end{figure}
    
For a foreground lensing galaxy at a redshift of $z_l$ = 0.89 having a virial mass of $7 \times 10^{11} \, M_{\odot}$ along with a background lensed galaxy at a redshift of $z_s$ = 1.51, we plot in Extended Data Fig.\,\ref{fig:nfwvar} the ratio $\sigma_\kappa(\boldsymbol{\xi})/\kappa_{\rm smooth}$ for which $\kappa_{\rm smooth}$ is the convergence for the adopted underlying smoothly-varying density profile.  The blue curve is for a $\psi$DM halo having a NFW global density profile as was used for making Figs.\,\ref{fig:kappa}--\ref{fig:qso2}, and the red curve for a $\psi$DM halo having a power-law (PL) global density profile as was used for making Fig.\,\ref{fig:hartley1}.  As can be seen, the fluctuations in column mass density of a $\psi$DM halo diminishes outwards from the halo centre.  The dashed vertical line indicates the Einstein radius for the model halos, around which the variation in $\sigma_\kappa/\kappa_{\rm smooth}$ with radius is quite similar for both halos.

   \begin{table*}[t!]
   \renewcommand{\tablename}{Extended Data Table}
    \begin{tabular}{p{2cm}cccp{4cm}}
      \toprule
      \toprule
Model &  S\'ersic index &Half-mass radius (kpc) &
        GRF Damping(\%) &  Baryonic fraction within Einstein radius (\%$M_h$) \\ \hline
      NFW & 4 & 1.1 & 20 & 3.40\\
      NFW & 4 & 1.0 & 50 & 7.50\\
      NFW & 4 & 0.4 & 80 & 13.3\\
      NFW & 1 & 2.0 & 20 & 3.70\\
      NFW & 1 & 1.3 & 50 & 7.70\\
      NFW & 1 & 1.2 & 80 & 13.5\\
            PL & 4 & 1.0 & 50 & 7.60\\
             \bottomrule
    \end{tabular}
    \caption{Relation between GRF damping and baryonic content for model $\psi$DM halos}
        \label{tab:sersic}
  \end{table*}

\subsection{Construction of $\psi$DM lenses}
To create the GRF capturing to the pervasive fuctuations in the column mass density of $\psi$DM halos about the local mean, we used the \textit{powerbox} package\cite{murray} in Python.  %These Gaussian profiles are assigned random amplitudes with a mean of 0 and, initially, a standard deviation of 1.  
The GRF has a mean of zero, and a standard deviation that varies with radius according to Eq.\,\ref{EM7} for a NFW profile -- and as plotted in Extended Data Fig.\,\ref{fig:nfwvar} (blue curve) for the NFW having parameters as listed in Extended Data Table\,\ref{tab:profiles}.
%To reflect the manner by which the fluctuation in column density changes with radius, we multiply the GRF with the standard deviation of the column density fluctuations, $\sigma_\Sigma(\boldsymbol{\xi})$, using Eq.\,\ref{EM7}.
Following standard practise \cite{dalal2021,kawai}, we adopted for each fluctuation (i.e., at each grid point) a Gaussian profile in column mass density with a full width at $\sigma$ of $\lambda_{\rm dB}$. In this way, the characteristic scale of fluctuations in the 3-D density field is preserved when projected onto 2-D, a good approximation when applying the central limit theorem to $\lambda_{\rm dB}$-sized regions that are independent in terms of phase for the interfering $\psi$DM waves and therefore of location.  An equivalent situation is temperature anisotropies in the Cosmic Microwave Background, whereby the power spectrum of the projected 2-D temperature fluctuations reflect that of the 3-D temperature fluctuations.

To construct the model $\psi$DM lenses used in Figs.\,\ref{fig:kappa}-\ref{fig:qso2}, we imprinted one or more of the GRF realisations thus generated at a selected $\lambda_{\rm dB}$ (for a given $m_\psi$) onto the NFW profile having parameters as listed in Extended Data Table\,\ref{tab:profiles}.  To simulate real galaxies, we added smoothly distributed baryons (stars and gas), the effect of which is to dampen fluctuations in the 3-D mass density and hence also fluctuations in the projected 2-D column mass density of a pure $\psi$DM halo.  As we are only interested in regions close to the critical curve, for simplicity we damp the GRF field by selected constant factors for the $\psi$DM halos used to make Figs.\,\ref{fig:qso}-\ref{fig:hartley1}.  For a given profile for the baryonic component, the damping thus imposed is related to the fractional baryonic mass within the Einstein radius.  Here, for illustration, we consider S\'ersic profiles for the baryonic component having S\'ersic indices of $n = 1$, corresponding to an exponential disk as is characteristic of spiral galaxies, and $n = 4$, corresponding to the de Vaucouleurs law as is characteristic of elliptical galaxies.  The half-mass radius of the baryonic component along with its fractional mass within the Einstein radius are listed in Extended Data Table\,\ref{tab:sersic} for the different damping factors considered in Figs.\,\ref{fig:qso}--\ref{fig:hartley1}.   

To avoid unnecessary computational effort given the similarity at which $\sigma_\kappa/\kappa_{\rm smooth}$ varies with radius around the Einstein radius for both NFW and PL profiles, to construct the suite of model $\psi$DM lenses employed in Fig.\,\ref{fig:hartley1}, we used the same GRF realisations but now imprinted onto a PL profile having parameters as listed in Extended Data Table\,\ref{tab:profiles}.

         \section{\textbf{\Large Gravitationally-Lensed Images}} 

Once a model for the surface mass density of the lensing galaxy is defined, the next step is to calculate deflection angles and magnifications imposed by gravitational lensing onto a background object (referred to in the formalism of gravitational lensing as the source).  The predicted positions and brightnesses of the resulting lensed images compared with those actually observed, if not in perfect agreement, leave position and brightness anomalies.

\subsection{\textbf{Deflection angles and magnifications}} \label{subsec:deflectionangle}

To accurately calculate deflection angles and magnifications imposed by gravitational lensing in the situation where the surface mass density of the lensing galaxy fluctuates on spatial scales characterised by $\lambda_{\rm dB}$, we need to adopt a pixel size that is much smaller than $\lambda_{\rm dB}$.  In our work, we adopt a pixel size of 1\,mas, which at a redshift of $z_l$ = 0.89 for the lensing galaxy corresponds to a physical size of 8\,pc, much smaller than range $\lambda_{\rm dB}  = 30$--210\,pc considered.  We generated multiple GRFs having a standard deviation that varies with projected radius as shown by the blue curve in Extended Data Fig.\,\ref{fig:nfwvar}.  Each realisation of the GRF spans a rectangular area of size 1000$\times$1000 pixels (8$\times$8 kpc).  Imprinting these GRFs onto the surface mass density of either the NFW (Figs.\,\ref{fig:kappa}--\ref{fig:qso2}) or PL (Fig.\,\ref{fig:hartley1}) halo having parameters as listed in Extended Data Table\,\ref{tab:profiles} to obtain different realisations of our model $\psi$DM lenses, we then computed deflection angles in each case for a lensed source at $z_s = 1.51$.  The computation of deflection angles requires significant computer time on clusters with large memory capacities, for which we used Python as well as Fortran with the OpenMP and MPI packages to take advantage of hybrid (inter-node and intra-node) parallel-processing techniques at the High Performance Supercomputing Cluster at the University of Hong Kong. The calculation of deflection angles was carried out on a pixel-by-pixel basis over a 2-D grid representing the lens plane.  Once the deflection angles are obtained for the entire grid, the magnifications were calculated in the standard manner by taking the determinant of the magnification tensor.  The lensing magnifications thus computed were used to plot critical curves (regions in the plane of the lensing galaxy corresponding to theoretically infinite magnification for background point sources) and caustics (mappings of the critical curves onto the source plane, thereby defining the locus of source positions that appear as images on the critical curve) for the gravitationally-lensed system.

    \begin{figure}[hbt!] %\vspace*{-4cm}
      \renewcommand{\figurename}{Extended Data Figure}
        \hspace{0cm}
      \includegraphics[width=7cm,height=7cm]{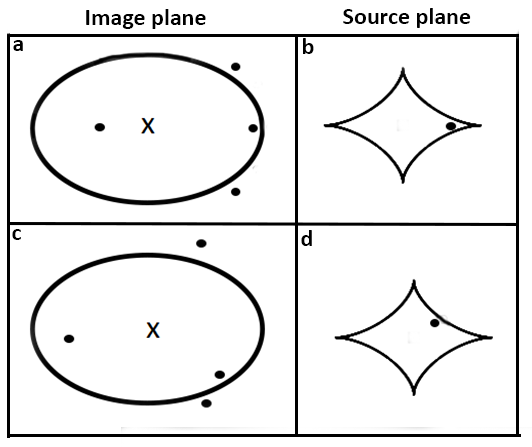}
      \caption{\textbf{Cusp and fold configurations}.  In the image plane, dots indicate multiply-lensed image positions of a compact background source.  Four images are distributed around an elliptical critical curve corresponding to the Einstein ring of the foreground lensing galaxy; a fifth image located near the lens centre as indicated by a $\times$ is not plotted as it is both demagnified and usually undetectable against the lensing galaxy.  In the source plane, a solitary dot indicates the position of the source located near a diamond-shaped caustic, which encloses the region that gives rise to five lensed images; if located along the caustic, two of these lensed images merge to appear at the critical curve.  \textbf{a--b,} Source located near a cusp of the caustic, giving rise to an image configuration whereby the three most closely-separated images are the most highly magnified.   \textbf{c--d,} Source located near a fold of the caustic, giving rise an image configuration whereby the two most closely-separated images are the most highly magnified.}
      \label{fig:caustics}
    \end{figure}
    
     \begin{figure}[hbt!] %\vspace*{-4cm}
       % \hspace{-1cm}
       \renewcommand{\figurename}{Extended Data Figure}
      \includegraphics[width=8cm,height=7cm]{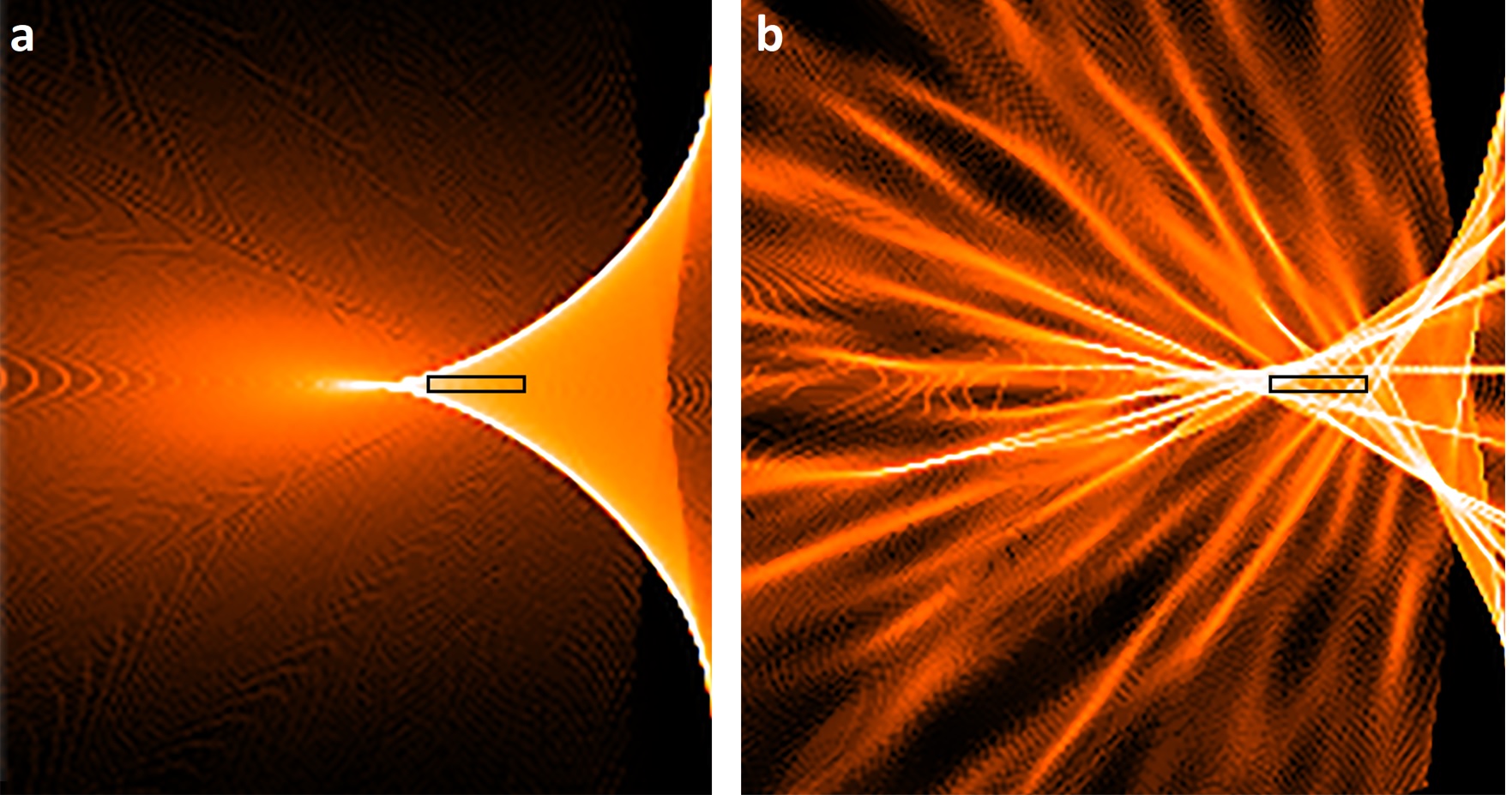}
        \caption{\textbf{Caustics}.  Close-up around a cusp of the caustic for a $\varrho$DM versus $\psi$DM halo that both have the same global profile (see Extended Data Table 1).   Colours indicate lensing magnification (brighter for higher magnification) imposed onto the brightest of the multiply-lensed images depending on where the source is located.   \textbf{a,} The simple and smooth caustic (white loci) of a $\varrho$DM halo, for which the fine striations are computational artefacts.   \textbf{b,} The complex caustic of a $\psi$DM halo with many branching micro-caustics, for which the finest striations are again computational artefacts.   Black rectangles bound the same region in the source plane near a cusp of the caustic, within which we placed a source at seventy-five different locations to mimic $\psi$DM halos created using seventy-five different GRF realisations (to save computation effort and time) for computing the positional and brightness anomalies in Fig\,\ref{fig:qso2}.}
      \label{fig:imgcfg}
    \end{figure}
    
  \begin{figure}[hbt!] %\vspace*{-4cm}
      \renewcommand{\figurename}{Extended Data Figure}
        \vspace{-0.8cm}
      \includegraphics[width=8cm,height=10cm]{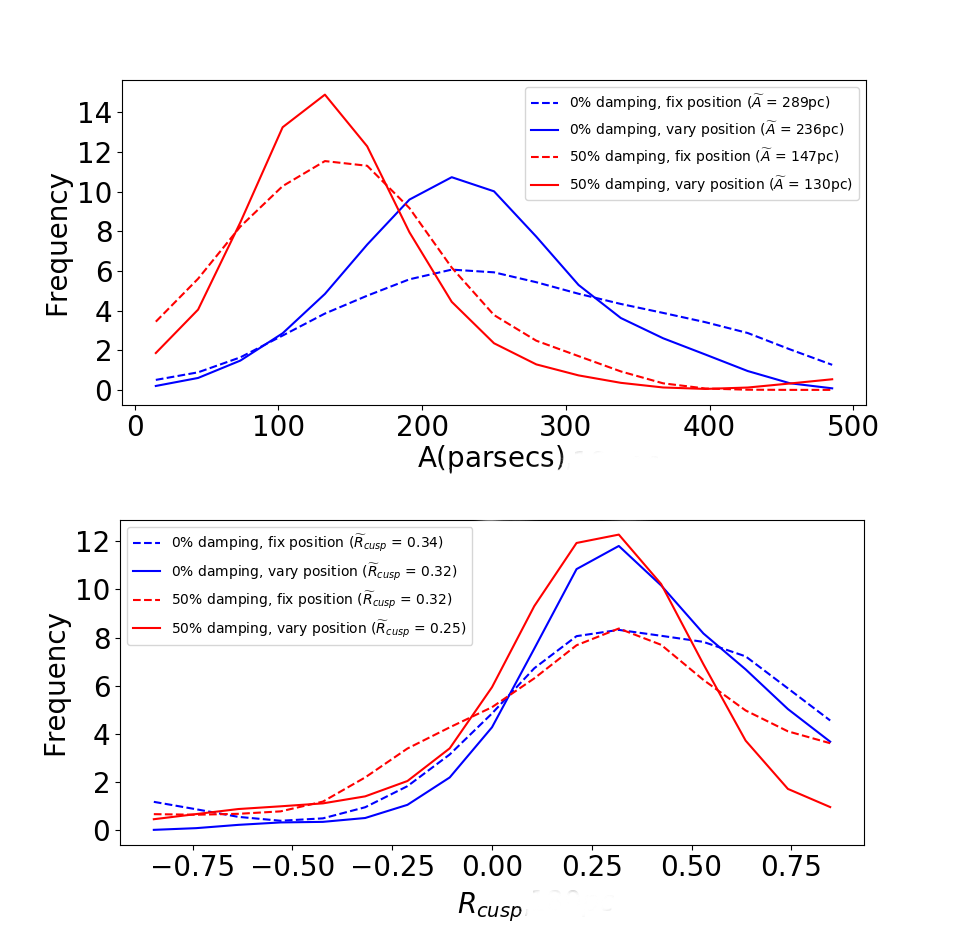}
      \caption{\textbf{Position and brightness anomalies computed using two approaches.} Probability distribution in \textbf{a,} position and \textbf{b,} brightness anomalies generated by: (i) imprinting 75 different GRF realisations having $\lambda_{\rm dB} = 180 \rm \, pc$ onto the $\varrho$DM lens of Fig.\,\ref{fig:qso} to create a suite of $\psi$DM lenses, and then generating multiply-lensed images for each $\psi$DM lens by placing the source at a fixed location (dashed curves); versus (ii) imprinting a single GRF realisation having also $\lambda_{\rm dB} = 180 \rm \, pc$ onto the $\varrho$DM halo, and generating multiply-lensed images by placing the source at seventy-five different locations within the black rectangle shown in Extended Data Fig.\,\ref{fig:caustics} (solid curves).  Blue curves are for pure $\psi$DM halos, whereas red curves are for $\psi$DM halos having density modulations damped by 50\% to include a baryonic component.  Although having similar medians, the second approach yields slightly elevated tails toward higher values.
      %The probability distributions computed using the two approaches have similar medians, but those based on a fixed source lensed by the 75 different GRF realisations of the model $\psi$DM lens have slightly elevated tails toward higher values.}
      }
      \label{fig:fig2test}
      \clearpage
    \end{figure}

\subsection{\textbf{Multiply-Lensed Image Configurations}} \label{subsec:images}

Extended Data Fig.\,\ref{fig:caustics} shows a schematic of diamond-shaped caustics (right column) typical of an elliptical lens, and the corresponding critical curves (left column).  As mentioned earlier, a source located on the caustic will produce an image on the critical curve.  By comparison, a source located near the cusp of a caustic as in Extended Data Fig.\,\ref{fig:caustics}$b$ will produce the configuration shown in Extended Data Fig.\,\ref{fig:caustics}$a$: (i) a highly-magnified triplet straddling the critical curve; (ii) a more weakly magnified image on the opposing side of the lensing galaxy that lies well within the critical curve; (ii) a demagnified image near the lens centre that is oftentimes undetectable (and therefore not shown).  This lensing configuration was used to generate the multiply-lensed images shown in Fig.\,\ref{fig:qso} and is the lensing configuration considered in making Fig.\,\ref{fig:qso2}.  On the other hand, a source located near the fold of a caustic as in Extended Data Fig.\,\ref{fig:caustics}$d$ will produce the configuration shown in Extended Data Fig.\,\ref{fig:caustics}$c$: (i) two highly-magnified and closely-separated images straddling the critical curve; (ii) two other more weakly-magnified and well-separated images lying inside and outside the critical curve respectively; and (iii) a demagnified image near the lens centre (again not shown).  
%which in the case of the CDM halo comprise (referring again to Methods Fig.\,\ref{fig:imgcfg}): (i) a highly-magnified triplet straddling the critical curve; (ii) a more weakly magnified image on the opposing side of the lensing galaxy that lies well within the critical curve (lying beyond the field shown); (ii) a demagnified image near the lens centre that is oftentimes undetectable (also beyond the field shown).  Owing to their high magnifications near the critical curve, only the triplet images are usually considered in studies of brightness and position anomalies.  

Extended Data Fig.\,\ref{fig:imgcfg}$a$ shows a portion of the caustic corresponding to the critical curve plotted in Fig.\,\ref{fig:qso}$a$ for a model $\varrho$DM lens.  By comparison, Extended Data Fig.\,\ref{fig:imgcfg}$b$ shows a portion of the caustic corresponding to the critical curve plotted in Fig.\,\ref{fig:qso}$d$ for a model $\psi$DM lens having the same global density profile as the model $\varrho$DM lens.  The caustic of the $\psi$DM lens is much more complex than that of a $\varrho$DM lens, reflecting the complex perturbations in the critical curve caused by random fluctuations in the surface mass density of the $\psi$DM lens.

 \subsection{\textbf{Brightness and Position Anomalies}} \label{subsec:Anomalies}
To mimic compact features such as quasi-stellar objects (seen in the optical or infrared) or jets from active supermassive black holes (commonly seen at radio wavelengths) in lensed background galaxies, we adopt Gaussian profiles for these feature (as is common practice in initial source modelling) with a full width at $\sigma$ of 2.5 pc.  Placed at a redshift $z_s$ = 1.51, the intrinsic angular size of the source is therefore 0.3\,mas.

As lensing preserves surface brightness, the brightness of each lensed image can be derived by calculating the ratio of the image area to the original source area, or by simply taking the value of the magnification at the pixel coordinate corresponding to the image centroid (for small lensed images).   As we simulate small sources, our lensed images are also small (verified through visual inspection of individual cases) and so for simplicity we use the latter approach.   In rare cases, however, it is possible for more than the number of images predicted in Extended Data Fig.\,\ref{fig:caustics} to form owing to the complex caustics -- comprising a large number of micro-caustics as evident in Extended Data Fig.\,\ref{fig:imgcfg}$b$ -- of $\psi$DM halos.   Many of these `extra' images are very faint, so in some cases it is possible to appropriately select just the brightest images in accordance with the predictions of Extended Data Fig.\,\ref{fig:caustics}, but when it is not possible to do so, we simply excluded source positions that gave rise to this situation and chose alternative source positions.

%The absolute brightnesses of the images can be obtained by calculating the ratio of the image area to the original source area, or by simply taking the magnification values at the coordinates of the central pixel of the image (for small sources).  Since we simulate small sources, our lensed images are also small (verified by going through individual cases) and so for simplicity we use the latter approach.  In a few rare cases, however, it is possible for a larger number of images to be created (4-6 as opposed to 3 for a cusp configuration)\cite{chan} as a result of the source being near a large number of microcaustics created by the DM fluctuations (see Fig 1b).  Most of the ‘extra’ images are very faint, so in some cases it is possible to still take the fluxes of the brightest images, but in situations where it is not possible to do so, we decide to exclude the source positions that give rise to this situation.

Because random fluctuations in the surface mass density fields of $\psi$DM halos are inherently indeterministic, we need to generate multiple GRF realisations all having the same $\lambda_{\rm dB}$ (corresponding to a particular halo and boson mass as defined by Eq.\,\ref{eq:m_psi1}) to capture the full range of deflections that such a halo can make in both the positions and brightnesses of multiply-lensed images.  While we generated seventy-five different GRF realisations all having $\lambda_{\rm dB} = 180 \rm \, pc$ to explore the range of deflections possible in Fig.\,\ref{fig:hartley1}, repeating this exercise for each of the $\lambda_{\rm dB}$ considered in Fig\,\ref{fig:qso2} along with the calculation of deflection angles for each the corresponding GRF realisation of the model $\psi$DM lens is too prohibitively expensive computationally.  Instead, to make Fig\,\ref{fig:qso2},
%rather than forward lensing a fixed source through multiple realisations of a model $\psi$DM lens having other choices of $\lambda_{\rm dB}$, 
we generated a single GRF having a particular $\lambda_{\rm dB}$ and varied the position of the source over a small region near the cusp of a caustic enclosed by the black rectangle in Extended Data Fig.\,\ref{fig:imgcfg}.  This practise is commonly adopted in the literature\cite{chan} to save computational time, and assumed implicitly to give similar results as independent GRF realisations for the model $\psi$DM lens.  Here, we show explicitly the close equivalence of these two methods. Extended Data Fig.\,\ref{fig:fig2test} shows the distribution in position (panel $a$) and brightness (panel $b$) anomaly generated by seventy-five different GRF realisations of the model $\psi$DM lens (having $\lambda_{\rm dB} = 180 \rm \, pc$) and a fixed position for the lensed source, to be compared to a single GRF realisation of the model $\psi$DM lens and varying the position of the source over seventy-five different locations in the region enclosed by the black rectangle in Extended Data Fig.\,\ref{fig:imgcfg}.  The distributions in both position and brightness anomalies computed using the two approaches have similar medians, but those based on a fixed source lensed by the seventy-five different GRF realisations of the model $\psi$DM lens have slightly elevated tails toward higher values in both position and brightness anomalies.

\section{\textbf{The system HS\,0810+2554}}
\subsection{\textbf{$\varrho$DM Lens Model}}
The system HS\,0810+2554 comprises a foreground early-type galaxy\cite{hartley} at a redshift of $z_l = 0.89$ that gravitationally lenses different components of a background galaxy at a redshift of $z_s = 1.51$, generating four detectable images of each component.  These components comprise a quasi-stellar object (QSO) at optical/near-IR wavelengths, for which the individual lensed images are spatially unresolved even in observations\cite{castles} with the Hubble Space Telescope (HST), and a pair of jets at radio wavelengths for which each jet is spatially resolved in observations\cite{hartley} using the European VLBI network that employs the method of Very Long Base Interferometry (VLBI).  Owing to present uncertainties in registering optical and radio reference frames that amount to a few 10\,mas, both sets of quadruply-lensed images at optical and radio wavelengths cannot be used simultaneously to inform lens models.  Because the images at radio wavelengths provide eight positional constraints whereas those at optical wavelengths provide only four, along with the much higher precision at which the image positions of the radio jets have been measured compared with those for the optical QSO, we used the positions of the quadruply-lensed pair of radio jets for informing lens models just like Ref.\,40. While it also is possible to use the brightness of the radio jets measured individually as constraints, like Ref.\,40 we chose not to do so as the brightnesses of the radio jets may not have been fully recovered owing to the lack of sufficiently short baselines in the EVN observations.  The ability of the lens model thus constructed to reproduce the observed brightnesses of the multiply-lensed images in HS\,0810+2554, informed only by the positions of these images, thus provides a measure of its reliability.

Ref.\,40 constructed a $\varrho$DM lens model for HS\,0810+2554 using a single isothermal ellipse (SIE) with and without external shear, finding that their lens model without external shear provided a somewhat better fit albeit still leaving positional anomalies among images of the radio jets. To make a broader assessment of how well $\varrho$DM lens models can reproduce the multiply-lensed images seen in HS\,0810+2554, we used the software algorithm \textit{glafic}\cite{glafic}. This algorithm permits various analytical profiles for galaxy halos (e.g., NFW, PL, SIE, and Jaffe profiles), along with additional components to mimic external shear as well as higher-order perturbations. To find the best-fit model, we tried various analytical profiles with all parameters including their centers set free, and allowed for external shear; in each case, the best-fit model was evaluated by minimising the reduced-$\chi^2$ based on differences between the predicted versus observed image positions of the radio jets. In this way, we found an elliptical PL profile with no external shear, the parameters for which are listed in Extended Data Table\,\ref{tab:profiles}, to provide the best fit.  

\begin{figure}[hbt!] %\vspace*{-4cm}
      \renewcommand{\figurename}{Extended Data Figure}
        \hspace{0cm}
      \includegraphics[width=8cm,height=6cm]{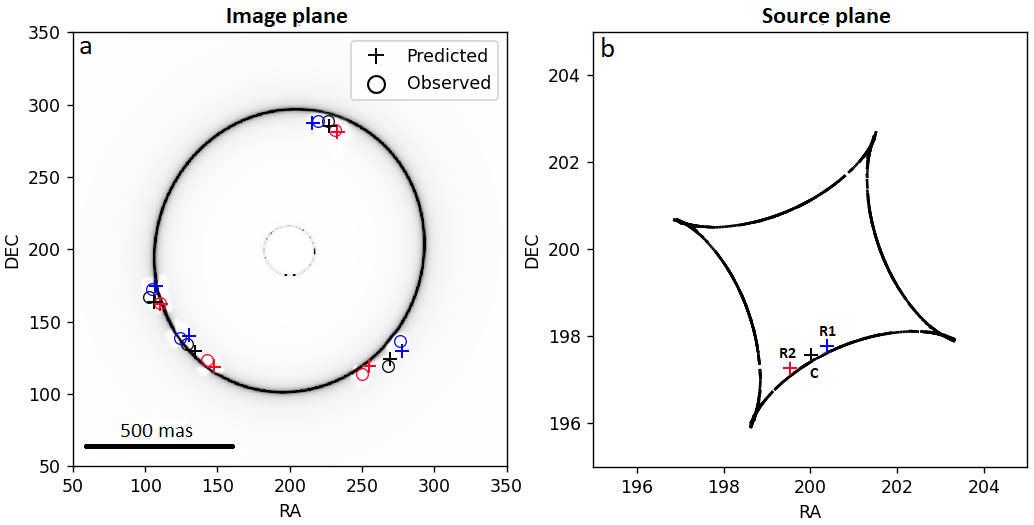}
      \caption{{\bf Predictions of best-fit $\varrho$DM lens model for HS\,0810+2554.}  \textbf{a,} Image positions of the quadruply-lensed pair of radio jets (red and blue crosses) and optical QSO (black crosses) predicted by our best-fit $\varrho$DM lens model.  Their observed image positions are indicated by the corresponding circles. Large black ellipse is the Einstein ring of our lens model at the redshift of the lensed source.  \textbf{b,}  Source positions for the pair of radio jets (labelled R1 and R2, respectively, for the blue and red crosses) and optical QSO (labelled C).  The QSO is assumed to correspond to the core of the radio jets, and therefore lie along a line connecting these jets.  Our best-fit lens model requires the core to be closer to radio jet R1 than R2, and all to be located near the caustic.}
      \label{fig:glafic}
    \end{figure}

Fig.\,\ref{fig:glafic}$a$ shows the image positions of the radio jets predicted by our best-fit $\varrho$DM lens model (crosses) versus those observed (ellipses), along with the Einstein ring for a lensed source at $z_s = 1.51$ (repeated from Fig.\,\ref{fig:hartley1}, which provides a magnified view, for ease of comparison).  Fig.\,\ref{fig:glafic}$b$ shows the corresponding model-predicted source positions of the pair of radio jets, along with the caustics of our best-fit $\varrho$DM lens model.   As can be seen, even our best-fit $\varrho$DM lens model leaves position anomalies among the multiply-lensed images. The magnitude of this position anomaly, $A$, for our best-fit $\varrho$DM lens model spans the range 15--26\,mas, similar to that reported by Ref.\,40.

\begin{figure*}[htb!]
\renewcommand{\figurename}{Extended Data Figure} 
\vspace*{0cm}
\centering
\hspace*{-1cm}
\includegraphics[width=1.1\textwidth,height=15cm]{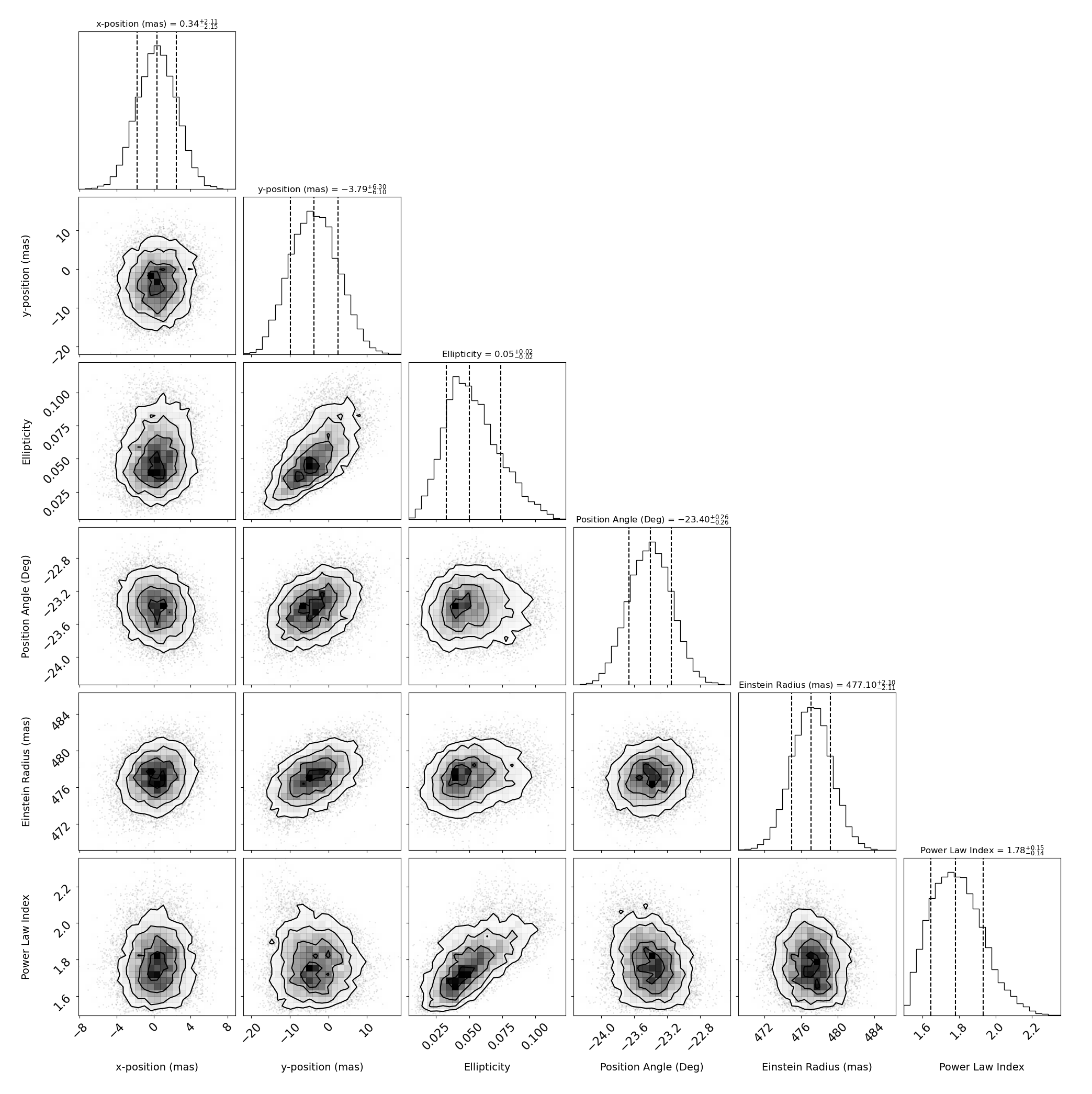}
\caption{{\bf Posterior probability distribution function for the parameters of our best-fit $\varrho$DM lens model for HS\,0810+2554.}  Histograms showing distribution in parameter values for each of the six parameters describing the PL model derived from a MCMC analysis.   Each parameter exhibits a distribution in values that can be closely described by a Gaussian having a mean and $\pm 1\sigma$ ranges as indicated by the dashed lines.  Contour plots with contour levels plotted at the 1$\sigma$, 2$\sigma$, and 3$\sigma$ indicate the correlation between pairs of model parameters.  Some of these parameters, like those between the ellipticity and y-position of the lens centre, exhibit strong correlations and therefore degeneracies.  The distributions in parameter values were used to estimate uncertainties in the predicted positions of the multiply-lensed images in HS\,0810+2554 as indicated by the crosses in Fig.\,\ref{fig:hartley1}.}
%Predicted (histograms) versus observed (dashed lines) brightness anomalies for the quadruply-lensed optical core treated as either a cusp ({\bf a}) or a fold ({\bf b}) configuration.   The brightness anomalies predicted by our best-fit CDM lens model lie within the narrow histograms (pink), which were generated by introducing small jitters to the position of the optical core twenty times to mimic tweaking of the best-fit CDM lens model parameters in different lens constructions.   The brightness anomaly predicted by the corresponding set of thirty-two $\psi$DM lens models (see caption for Fig.\,\ref{fig:hartley1}) are indicated by the broad histograms (blue).   Whereas the brightness anomalies predicted by the CDM lens model are in clear tension with those measured, the brightness anomalies predicted by the corresponding $\psi$DM lens models span a range that encompasses, and furthermore when considered as a cusp configuration peaks at or close to, the measured brightness anomalies.}
 \label{fig:mcmc}
\end{figure*}

As explained above, the best-fit model $\varrho$DM lens that we constructed is informed by the image positions of the quadruply-lensed pair of radio jets, which have 1$\sigma$ measurement uncertainties in their positions of typically 1\,mas (and no larger than 4\,mas) as reported by Ref.\,40. These position uncertainties translate to tolerances in the individual parameters of the model $\varrho$DM lens, that in turn translate into uncertainties in the model-predicted image positions of each radio jet.  To estimate the uncertainties in the model-predicted image positions, we employ a Markov chain Monte Carlo (MCMC) method using \textit{glafic}\cite{glafic} by providing Gaussian priors centered on the best-fit model parameters: 2-D position of the centre, ellipticity, position angle, power law-index, and Einstein radius of the lens.  Out of 50,000 realisations with a 0.214 acceptance rate, we were left with 10,700 realisations.  To exclude likely unphysical outliers among acceptable realisations, we selected only those having parameters that are simultaneously within the top 68\% (1$\sigma$) of their respective probability distributions, thus leaving 1728 possible combinations of lens model parameters that satisfy the image positional constraints given their measurement uncertainties. Extended Data Fig.\,\ref{fig:mcmc} shows the joint posterior probability distribution for different pairs of model parameters (contours), as well as the joint posterior probability distribution for each model parameter separately (histograms).
%We show the posterior probability distribution function plots in Supplementary Fig.5, with dashed black lines indicating $\pm \sigma$ boundaries for each lens model parameter. 
The tolerances thus computed in the model $\varrho$DM lens give rise to uncertainties in the image positions of the individual radio jets with standard deviations, $\sigma_{\rm tol}$, ranging from 1.2--3.2\,mas, comparable with the measurement uncertainties for the actual image positions of the individual radio jets.  The crosses plotted in Fig.\,\ref{fig:hartley1} and Extended Data Fig.\,\ref{fig:glafic}$a$ have arm lengths of $\pm 3\sigma_{\rm tol}$.  The position anomalies left by our best-fit $\varrho$DM lens model are therefore significantly larger than the measured and model-predicted positional uncertainties combined.

%As mentioned in both the main text and Methods, the accuracy at which optical and radio reference frames can be registered is a few 10\,mas.  Our smooth lens model for HS\,0810+2554 was constructed using the image positions of all the quadruply-lensed pairs of radio jets as constraints.  
Registering the optical and radio reference frames using the coordinates for both images as reported (i.e., without introducing any arbitrary shift between these frames), we find that the source position that best reproduces the measured positions of the quadruply-lensed images of the optical QSO -- if we require it to lie between and along a line connecting the source positions of the radio jets, as would be expected -- is about one-third closer to one jet than the other, as shown by the black cross in Extended Data Fig.\,\ref{fig:glafic}$b$.  This configuration is commonly seen for core-jet systems, where the optical QSO comprises the core, owing to Doppler boosting that brightens one jet but dims the opposing jet.  The corresponding quadruply-lensed image positions of this QSO is shown by the black crosses in Extended Data Fig.\,\ref{fig:glafic}$a$, for which their arm lengths are $\pm 3\sigma_{\rm tol}$ as before.

%Fig.\,\ref{fig:glafic} shows the a) image plane and b) source plane for the CDM power law model best-fit, and the lensed images and source positions corresponding to the two radio jets.

\subsection{\textbf{$\psi$DM Lens Model}}

We constructed a suite of $\psi$DM lens models for HS\,0810+2554 by imprinting 75 different GRF realisations onto our best-fit $\varrho$DM lens model for this system having parameters as listed in Extended Data Table\,\ref{tab:profiles}.  The manner by which the standard deviation of the GRF changes with radius from the centre of a PL halo is shown in Extended Data Fig.\,\ref{fig:nfwvar} (red curve); near the Einstein radius, the trend is similar to that for a NFW profile (blue curve) having an identical Einstein radius.  Thus, to avoid unnecessary computational effort and time, we used the same GRF realisations as constructed specifically for the NFW profile.  The standard deviation of these GRFs was damped by 50\% to allow for a corresponding fractional baryonic mass of 7.6\% $M_h$ within the Einstein radius, as would be provided by an elliptical galaxy described by a S\'ersic profile with an index of $n = 4$, half-mass radius of 1\,kpc, and an ellipticity of 0.2 like that of its $\psi$DM halo (see Extended Data Table\,\ref{tab:sersic}).

Based on this suite of model $\psi$DM lenses and the source positions for the radio jets and optical QSO as inferred from our best-fit $\varrho$DM lens model (shown in Extended Data Fig.\,\ref{fig:glafic}$b$), the image positions predicted for the radio jets and optical QSO are shown by the dots in Fig.\,\ref{fig:hartley1}$a$--$d$.  The uncertainty in their predicted image positions is similar to that for the predicted image positions of the model $\varrho$DM lens as indicated by the cross lengths.  As can be seen, unlike for our best-fit model $\varrho$DM lens, the image position predicted by a given realisation of the model $\psi$DM lens (having the same global density profile as the $\varrho$DM lens) can agree with its observed position.  Extended Data Fig.\,7 shows the level of positional agreement for the individual lensed images based on just a single realization of the model $\psi$DM lens.

At this point, we wish to emphasise once again (as mentioned in the main text) that the success or otherwise of our $\psi$DM lens model should not be evaluated based on whether or how well any single realisation of the model $\psi$DM lens  reproduces both the positions and brightnesses of all the lens images simultaneously.  Owing to the random and independent nature of surface density fluctuation in a $\psi$DM halo along different sightlines, any number of realisations can only comprise a small subset of all possible outcomes.  Instead, the success of our $\psi$DM lens model should be evaluated based on whether the suite of different realizations explored can produce the level of brightness and positional anomalies seen, together (as is possible in the case of HS\,0810$+$2554) with the battery of other tests as described in the SI.

%Extended Data Fig.\,\ref{fig:bestfit} shows the best-fit realization out of the 75 $\psi$DM lens models. The best-fit was decided by calculating the position anomaly between the observed and $\psi$DM model predicted image positions and finding the minimum position anomaly on average for all 12 images.}

\begin{figure*}[htb!]
\renewcommand{\figurename}{Extended Data Figure} 
\vspace*{0cm}
\centering
\hspace*{-1cm}
\includegraphics[width=1\textwidth,height=12cm]{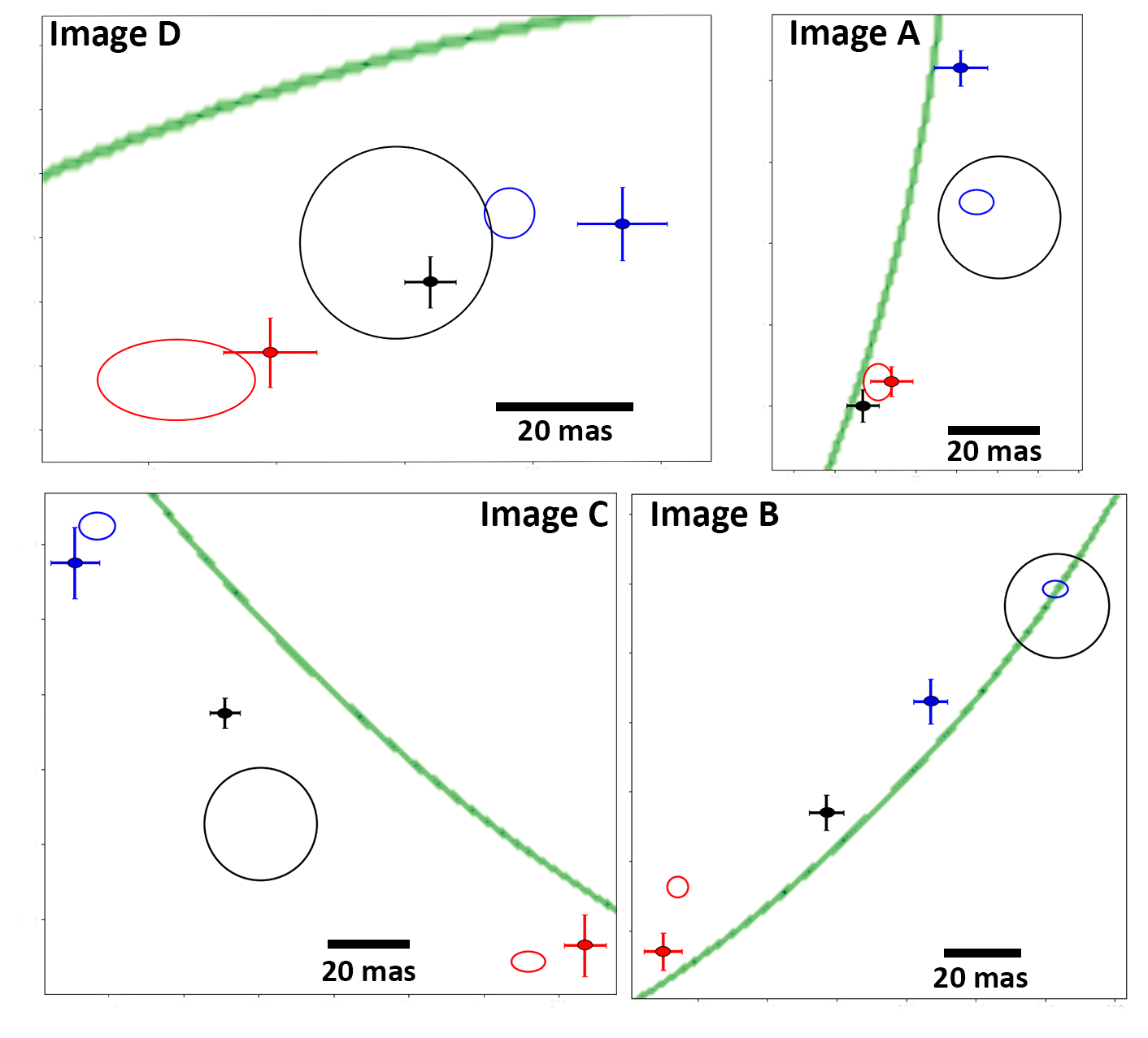}
\caption{A single realization of our model $\psi$DM lens that yields the best overall agreement between the predicted and observed positions of the multiply-lensed radio images (indicated by blue and red symbols) among the seventy-five different realizations explored.  We do not consider how well our model $\psi$DM lens reproduces the HST positions of the multiply-lensed QSO (indicated by black symbols) as there is an uncertainty of a few 10\,mas between the registration of the radio and optical reference frames.  As in Fig.\,\ref{fig:hartley1}, observed image positions are indicated by ellipses, each with a radius or semi-major and semi-minor axes of $3\sigma$ (where $\sigma$ is the measurement uncertainty) so as to encompass 99.7\% of all possible positions.  Dots indicate the positions of the quadruply-lensed images predicted by this particular realization of the model $\psi$DM lens, with arm lengths corresponding to $\pm 3 \sigma_{\rm tol}$ (for which $\sigma_{\rm tol}$ reflects tolerances in the inferred parameters of the model $\varrho$DM lens as described in Methods).}
\label{fig:bestfit}
\end{figure*}

    \begin{center}
\textbf{\large Data availability}
    \end{center}

The authors welcome requests to collaborate, and will share the data used in this study accordingly.

 \begin{center}
\textbf{\large Code availability}
    \end{center}
    
The authors welcome requests to collaborate, and will share the code used in this study accordingly.

  \begin{center}
\textbf{ \large Acknowledgements}
\end{center}

We would like to thank the referees (Lam Hui, Antonio Herrera-Martin and the anonymous) for helping improve the presentation of the paper immensely. J.L. acknowledges a grant from the Research Grants Council of Hong Kong through the General Research Fund 17304519, which also partially supports a graduate scholarship for A. A. T.J.B. is supported by the Spanish project grant PID2020-114035GB-100 (MINECO/AEI/FEDER, UE) as well as the General Research Fund 17304519. M.O. acknowledges the support by World Premier International Research Center Initiative (WPI Initiative), MEXT, Japan, and JSPS KAKENHI Grant Numbers JP20H04725, JP20H00181, JP20H05856 and JP18K03693. J.M.D. acknowledges the support of projects PGC2018-101814-B-100, and MDM-2017-0765. R.E. acknowledges the support by the Institute for Theory and Computation at the Center for Astrophysics as well as grant numbers 21-atp21-0077, NSF AST-1816420 and HST-GO-16173.001-A. H.S. acknowledges funding support from the Jade Mountain Young Scholar Award No. NTU-111V1201-5, sponsored by the Ministry of Education, Taiwan, as well as by the National Science and Technology Council (NSTC) of Taiwan under Grant No. NSTC 111-2628-M-002-005-MY4 and the NTU Academic Research-Career Development Project under Grant No.NTU-CDP-111L7779. This work was carried out in part using the High Performance Computing (HPC) facilities at the University of Hong Kong.

    \begin{center}
\textbf{ \large Author contributions}
\end{center}
A.A. wrote original parallel computing code to perform $\psi$DM lensing simulations, performed analysis of results and contributed to interpretation of results. T.B. and J.Lim supervised the project and contributed to interpretation of results. J.Lim, A.A. and T.B. wrote the paper. J.D. contributed to lensing simulations and interpretation of results. G.S. contributed to analysis of Gaussian random fields and interpretation of results. M.O., E.L., R.E. contributed to constructing the mathematical framework of $\psi$DM density fluctuations. T.C., H.S. and M.Y. contributed to methodology for generating Gaussian random fields. J.Li contributed to lens modelling using \textit{glafic}. S.K.L. and J.Li carried out MCMC calculations.

    \begin{center}
\textbf{ \large Competing interests}
\end{center}
The authors declare no competing interests.\\

\newpage 

\begin{center}
  
\textbf{\huge Supplementary Information}

\end{center}

\noindent \textbf{Additional tests of $\psi$DM lens model for HS0810+2554.}  In the main text, we described three tests for assessing whether a $\psi$DM lens is responsible for generating the multiply-lensed images observed in HS\,0810+2554:\ the ability of such a lens to reproduce the position (Test\,1) and brightness (Test\,2) anomalies left by a $\varrho$DM lens having the same global density profile for which the results are shown in Fig.\,\ref{fig:qso2}, along with its ability to reproduce the observed image positions of the multiply-lensed sources (Test\,3) for which the results are shown in Fig.\,\ref{fig:hartley1}.  In the following, we describe three additional and independent tests that supplement the above tests.   \\

\begin{figure*}
     \renewcommand{\figurename}{Extended Data Figure} 
%     \begin{subfigure}[t!]{1\textwidth}
        \vspace{-5cm}
        \hspace{-3cm}
\includegraphics[width=22cm,height=17cm]{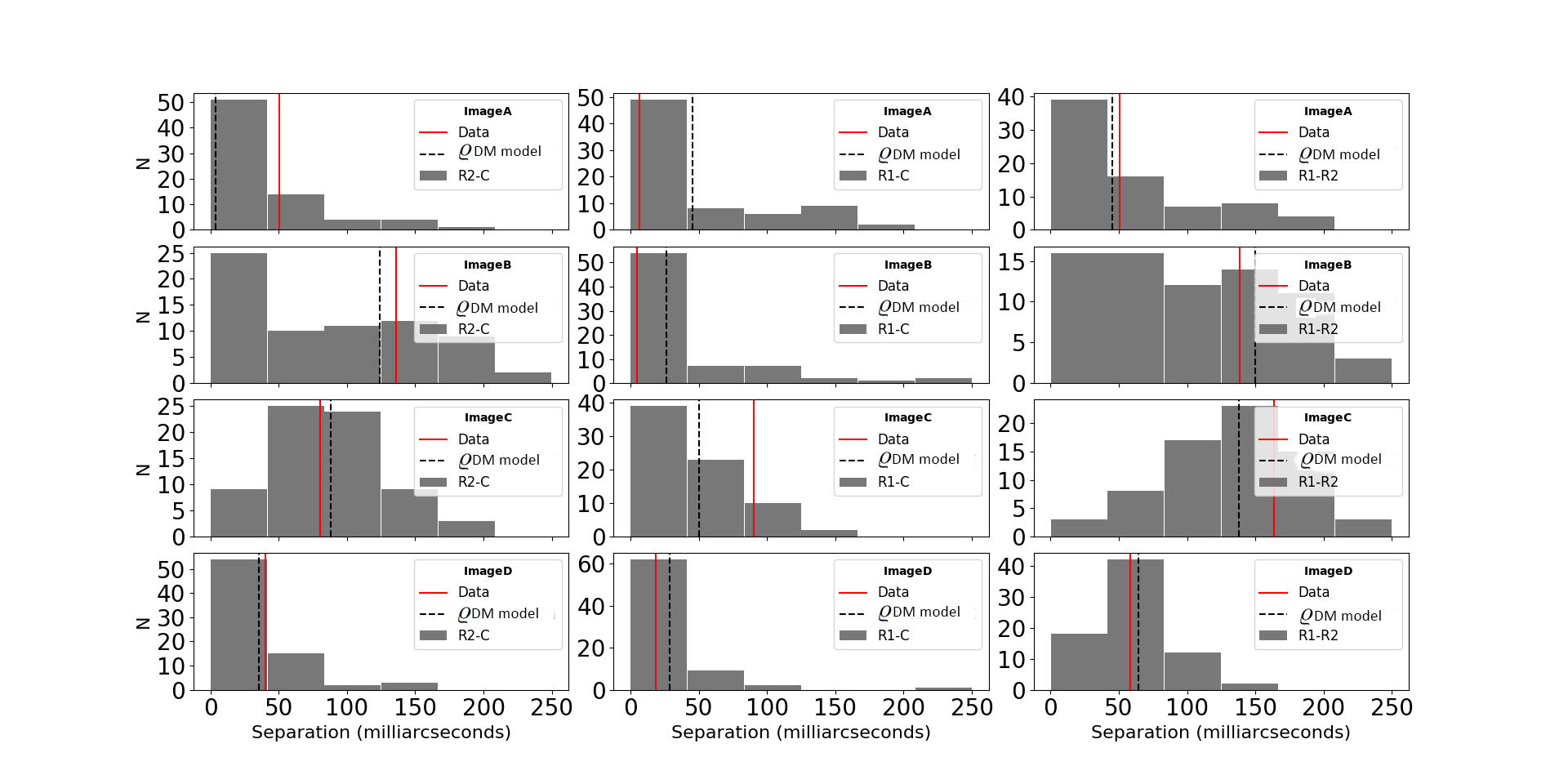}
        \caption{{\bf Pair separations (Test\,4).}  Histograms indicating angular separations between two of the lensed components in each of the four sets of multiply-lensed images in the system HS0810+2554 as predicted by 75 different realisations of our $\psi$DM lens model.  All combinations of pair separations -- between the radio jets R1-R2, as well as between the optical QSO and each radio jet R1-C and R2-C (see labelling in Extended Data Fig.\,\ref{fig:glafic}) -- are shown for each of the four sets of multiply-lensed images (labelled A-D in accordance with Fig.\,\ref{fig:hartley1}).  Dashed black lines indicate the pair separations predicted by our best-fit $\varrho$DM lens model, and red lines the actual observed pair separations.  Whereas the $\varrho$DM lens model leave anomalies in pair separations, the median pair separations predicted by the $\psi$DM lens model are in good or close agreement with those observed in the majority of cases.}
        %        %{\bf Histogram of image separation distances.} We plot the distribution of image separations (1 pixel $\sim$ 1 milliarcsecond) between lensed images of the 3 sources (2 radio jets and optical core) for Image A over 40 $\psi$DM realizations.  Image separations of observed images are shown as dashed lines for each case. Red, green and blue histograms correspond to separation distances between the images of Radio source 1 - Optical source, Radio source 1 - Radio source 2 \& Radio source 2 - Optical source respectively.}
%\subcaption{Image A}
\label{fig:pairsep1}
%     \end{subfigure}
     \end{figure*}

\noindent \textbf{Test 4: Pair separations.}  This test considers each image pair in HS\,0810+2554 together, by contrast with Test\,3 that considers each image separately.  Specifically, for each set of the quadruply-lensed images, we consider the separation between the pair of radio jets, referred to as the pair separation R1-R2 (see labelling in Extended Data Fig.\,\ref{fig:glafic}).  In addition, we also consider the separation between the optical QSO and each individual radio jet, referred to as R1-C and R2-C.
%As we have generated 40 diffferent GRF realizations to make Fig.3, we can compute the image pair separations using them.  In the source plane, we place the optical core 33\% closer to the radio jet marked in red on Methods Fig. 4.    That is, the image pairs considered comprise all possible combinations of the two radio jets (referred to as A and B) and the single optical QSO (referred to as C): A-B, A-C, and B-C, labelled in accordance with Supplementary Fig.\,\ref{}. 
Histograms in Extended Data Fig.\,\ref{fig:pairsep1} show the image pair separations for each of the four sets of multiply-lensed images predicted by our model $\psi$DM lens.  Like for the position and brightness anomalies (Fig.\,\ref{fig:qso2}), the model-predicted pair separations span a range owing to different patterns of random fluctuations in the surface mass density arising from the different realisations of the model $\psi$DM lens.  By comparison, the dashed black lines indicate the pair separations predicted by our best-fit model $\varrho$DM lens, and the red lines the actual observed pair separations.  Whereas the $\varrho$DM lens model leave anomalies in pair separations just like those in positions (Test\,1) and brightnesses (Test\,3), the pair separations predicted by the $\psi$DM lens model encompass the observed separations.  Moreover, the $\psi$DM lens model predicts a preferred pair separation in all four sets of multiply-lensed images such that, in the majority of cases, the median pair separations are in good or close agreement with those observed.   \\

\textbf{Test 5: Pair brightness ratios.}  Another important challenge for lens models is their ability to correctly predict the ratio in brightnesses of the observed lensed images, in particular when the relative brightnesses of lensed images are not used to constrain the lens model -- as is the case here.  In the case of the radio jets in HS\,0810+2554 as observed by Ref.\,40, there is a possibility that the brightness of the individual radio jets may not have been fully recovered owing to the lack of short baselines in the EVN observations.  Thus, in this particular instance, the pair brightness ratio test may not be as precise as the pair separation test (Test\,4) for the radio jets. 

\begin{figure*}[t!]
    \renewcommand{\figurename}{Extended Data Figure} 
        \vspace{-3cm}
        \hspace{-1.2cm}
\includegraphics[height=15cm,width=18cm]{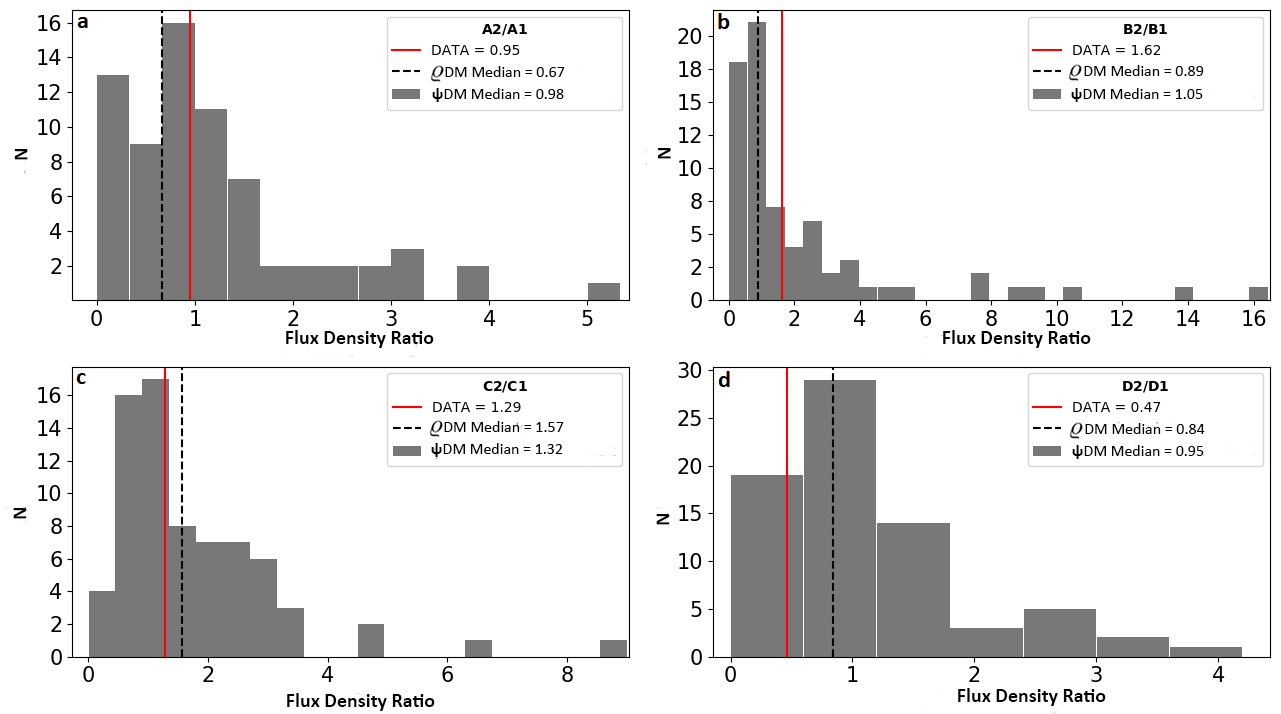}
      \caption{{\bf Brightness ratio between image radio jet pairs (Test 5).}  Histograms indicating brightness ratio between the pair of radio jets in each of the four sets of multiply-lensed images (labelled A-D in accordance with Fig.\,\ref{fig:hartley1}) in the system HS0810+2554 as predicted by 75 different realisations of our $\psi$DM lens model.  The brightness ratios predicted by our best-fit $\varrho$DM lens model is shown by the dashed black lines, and those actually observed by the red lines.  Whereas the smooth and $\psi$DM lens model are able to correctly predict the observed brightness ratios of the radio jets comparably well for images A and D, the prediction of the $\psi$DM lens model is superior to that of the $\varrho$DM lens model for image C and, especially, image B.}
      \label{fig:fluxanomaly1}
    \end{figure*}

Extended Data Fig.\,\ref{fig:fluxanomaly1} shows the ratios in flux densities between the two radio jets in each of the four sets of multiply-lensed images as are observed (red line), versus those predicted by our best-fit model $\varrho$DM lens (back dashed line) and the suite of model $\psi$DM lenses (histogram).   Both the $\varrho$DM and $\psi$DM lens models are able to correctly predict the observed brightness ratios of the radio jets comparably well for images A and D (see Fig.\,\ref{fig:hartley1} for how the four sets of multiply-lensed images are labelled), whereas the prediction of the $\psi$DM lens model is superior to that of the $\varrho$DM lens model for image C and, especially, image B.   \\

\begin{figure*}[htb!]
\renewcommand{\figurename}{Extended Data Figure} 
\vspace*{-3cm}
\centering
\hspace*{-2cm}
\includegraphics[width=1.2\textwidth,height=12cm]{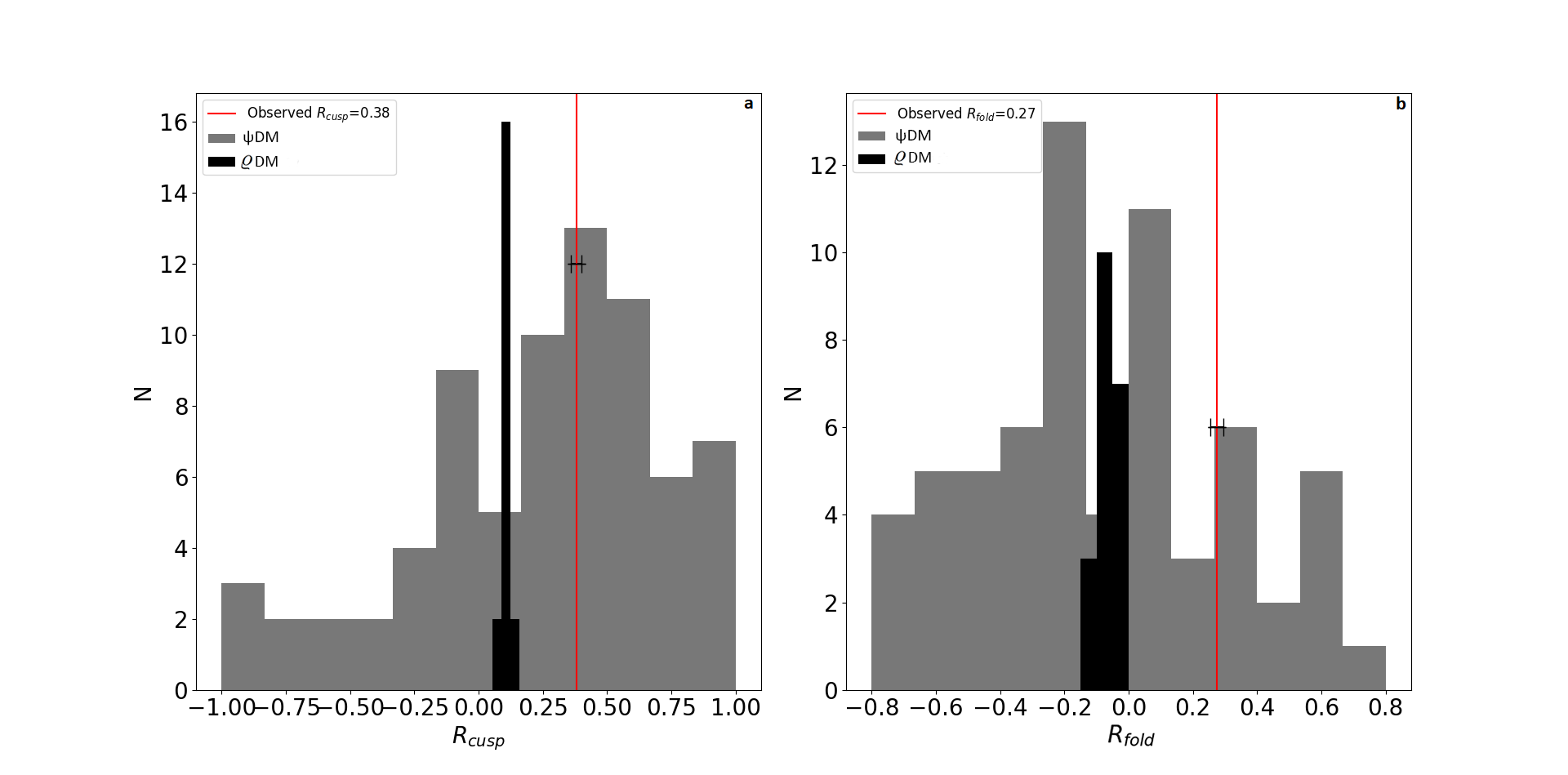}

\caption{{\bf Brightness anomalies for optical QSO (Test\,6).}   Grey histograms indicating brightness anomalies for the quadruply-lensed optical QSO in HS0810+2554 as predicted by 75 different realisations of our $\psi$DM lens model.  Narrow black histograms show the corresponding brightness anomalies predicted by our best-fit $\varrho$DM lens model, whereas red lines indicate the measured brightness anomaly.  {\bf a,} Treated as a cusp configuration, the brightness anomalies correspond to that for the three most-closely separated multiply-lensed images (Images A--C in Fig\,\ref{fig:hartley1}).  {\bf b,} Treated as a fold configuration, the brightness anomalies correspond to that of the two most-closely separated multiply-lensed images (Images A--B in Fig\,\ref{fig:hartley1}).  The distribution in the predicted brightness anomaly spans a range that encompasses, and when considered as a cusp configuration peaks at or close to, the measured brightness anomaly.  When treated as a fold configuration, the $\varrho$DM lens model does not correctly predict that the image inside the critical curve is actually dimmer than that outside the critical curve, a possibility allowed by the $\psi$DM lens model.
}
%Predicted (histograms) versus observed (dashed lines) brightness anomalies for the quadruply-lensed optical core treated as either a cusp ({\bf a}) or a fold ({\bf b}) configuration.   The brightness anomalies predicted by our best-fit CDM lens model lie within the narrow histograms (pink), which were generated by introducing small jitters to the position of the optical core twenty times to mimic tweaking of the best-fit CDM lens model parameters in different lens constructions.   The brightness anomaly predicted by the corresponding set of thirty-two $\psi$DM lens models (see caption for Fig.\,\ref{fig:hartley1}) are indicated by the broad histograms (blue).   Whereas the brightness anomalies predicted by the CDM lens model are in clear tension with those measured, the brightness anomalies predicted by the corresponding $\psi$DM lens models span a range that encompasses, and furthermore when considered as a cusp configuration peaks at or close to, the measured brightness anomalies.}
 \label{fig:fluxanomaly}
\end{figure*}

\noindent \textbf{Test 6: QSO Brightness anomalies.}  
%The brightnesses of the quadruply-lensed radio jets are not sufficiently well defined in the EVN image to test whether the $\psi$DM lens model also can correctly predict their relative brightnesses, 
Unlike the image brightnesses of the radio jets as measured in the EVN observation, the image brightnesses of the optical QSO are reliably measured in the HST observation.   As the overall geometry of the quadruply-lensed images in HS\,0810+2554 indicates a configuration between between a cusp and a fold (see Extended Data Fig.\,\ref{fig:glafic}), in Extended Data Fig.\,\ref{fig:fluxanomaly} we plot the measured brightness anomalies (red lines) of the QSO for both configurations.  The brightness anomaly for a cusp configuration has already been defined in the main text, whereas that for a fold configuration is defined as $R_{\rm fold} =\frac{\mu_1 + \mu_2}{|\mu_1| +|\mu_2|}$, where $\mu_1 > 0$ and $\mu_2 < 0$ correspond to the most closely separated and also the most highly magnified image pair.   The brightness anomaly predicted by our best-fit $\varrho$DM lens model is shown by the black histogram, which is generated by moving the source around such that its image positions change over the range $\sim$10--30\,mas (depending on a particular image, with the range corresponding approximately to the precision in registering the optical and radio reference frames).  Treated either as a cusp or a fold configuration, the measured brightness anomaly is in strong tension with that predicted by our best-fit $\varrho$DM lens model or, indeed, that of Ref.\,40.
%, corresponding to a value close to zero and lying within the narrow black histograms for the two different configurations.   

The brightness anomalies predicted by the $\psi$DM lens model are shown by the broad grey histograms.  Once again, there is a spread in the predicted brightness anomalies owing to different patterns of random fluctuations in the surface mass density arising from the different realisations of the $\psi$DM lens model.   The distribution is roughly symmetric about zero for the $R_{\rm fold}$ parameter, but has a positive skew for the $R_{cusp}$ parameter for the reasons mentioned in the main text.  Treated as either a cusp or a fold configuration, the distribution in the predicted brightness anomaly spans a range that encompasses, and furthermore when considered as a cusp configuration peaks at or close to, the measured brightness anomaly.   Note that, in the case of a cusp configuration, this test also evaluates whether the lens model correctly predicts that the brightness of the image within the critical curve is lower (in which case $R_{\rm cusp} > 0$) than the combined brightnesses of the two images outside the critical curve.  Likewise, for a fold configuration, this test evaluates whether the lens model correctly predicts that the brightness of the image within the critical curve is lower (in which case $R_{\rm fold} > 0$) than the brightness of the image outside the critical curve.  When treated as a fold configuration, the best-fit model $\varrho$DM lens does not correctly predict that the image inside the critical curve is actually dimmer than that outside the critical curve, a possibility allowed by the $\psi$DM lens model.
\\

All the aforementioned tests (six altogether) demonstrate the predictive power of our $\psi$DM lens model for reproducing the observed positions and brightnesses of the radio jets and optical QSO in HS0810+2554 -- even though the lens model is informed only by the observed positions of the radio jets.  By contrast, our best-fit $\varrho$DM lens model and, where subjected to the same tests, also the best-fit $\varrho$DM lens model of Ref.\,40 fail nearly all of these tests.  Satisfying all these tests simultaneously pose as critical challenges for $\varrho$DM lens models employing additional ingredients not contemplated here (e.g., sub-halos).

\end{document}